%% file: PDE-ODE_cascade.tex
\documentclass[12pt]{article}
\usepackage{amsmath,amsthm,amssymb}
\usepackage{graphicx,epstopdf,framed}
\usepackage{euscript,paralist}

\input{header}

\begin{document}
\renewcommand{\thefootnote}{\fnsymbol{footnote}}
\renewcommand{\thefootnote}{\fnsymbol{footnote}}
\newcommand{\footremember}[2]{%
   \footnote{#2}
    \newcounter{#1}
    \setcounter{#1}{\value{footnote}}%
}
\newcommand{\footrecall}[1]{%
    \footnotemark[\value{#1}]%
}
\makeatletter
\def\blfootnote{\gdef\@thefnmark{}\@footnotetext}
\makeatother

\begin{center}
{\Large \bf Compensating PDE actuator and sensor \\[1ex]
  dynamics using Sylvester equation}\\[2ex]
Vivek Natarajan
\blfootnote{This work was supported by the Industrial Research and Consultancy Centre at IIT Bombay via the seed grant 16IRCCSG004 and the Science and Engineering Research Board, DST India, via the grant ECR/2017/002583.}
\blfootnote{V. Natarajan (vivek.natarajan@iitb.ac.in) is with the Systems and Control Engineering Group, Indian Institute of Technology Bombay, Mumbai, India, 400076, Ph:+912225765385.}
\end{center}

\begin{abstract} {{\noindent
We consider the problem of stabilizing PDE-ODE cascade systems in which the input is applied to the PDE system whose output drives the ODE system. We also consider the dual problem of constructing an observer for ODE-PDE cascade systems in which the output of the ODE system drives the PDE system, whose output is measured. The PDE in these problems is stable and the ODE is unstable. While the ODE system models the plant in both the problems, the PDE system models the actuator in the stabilization problem and the sensor in the dual problem. In the literature, these problems have been solved for specific PDE models using the backstepping approach. In contrast, in the present work we consider these problems in an abstract framework by letting the PDE system be any regular linear system. Using a state transformation obtained by solving a Sylvester equation with unbounded operators, we first diagonalize the state operator corresponding to the cascade systems. We then solve the stabilization problem and the dual estimation problem, provided they are solvable, by solving certain finite-dimensional counterparts. We also derive necessary and sufficient conditions for verifying the solvability of these problems. We show that the controller which solves the stabilization problem is robust to certain unbounded perturbations. We illustrate our theory by designing a stabilizing controller for a PDE-ODE cascade in which the PDE is a 1D diffusion equation and an observer for a ODE-PDE cascade in which the PDE is a 1D wave equation.}} \vspace{-2mm}
\end{abstract}

 \vspace{-2mm}

\section{Introduction} \label{sec1}
\setcounter{equation}{0} 
\vspace{-1mm}

\ \ \ Consider an unstable finite-dimensional linear plant. Suppose that this plant is driven via an actuator with stable PDE (partial differential equation) dynamics which is not influenced by the plant dynamics (i.e. the actuator is sufficiently strong). Then the actuator-plant model is a cascade interconnection of a PDE system driven by an input and an ODE (ordinary differential equation) system driven by the output of the PDE system.  Similarly, suppose that the plant output is measured using a sensor with stable PDE dynamics which does not influence the plant dynamics. Then the plant-sensor model is a cascade interconnection of a ODE system whose output drives the PDE system, whose output is in-turn measured. In this paper, we address the problem of designing state and output feedback control laws for stabilizing the former interconnection and the problem of designing an observer for the latter interconnection.

Motivated by applications in chemical process control, combustion systems, traffic flow and water channel flow, the above stabilization and estimation problems have been solved for many specific one-dimensional PDE models by Krstic and coauthors using the backstepping method, see \cite{LiKr:2010}, \cite{Kri:2009}, \cite{Kri:2009a}, \cite{Kri:2010}, \cite{KrSm:2008}, \cite{SuKr:2010}. In \cite{KrSm:2008}, the actuator dynamics and sensor dynamics, which are pure delays, are compensated by first modeling them using first-order hyperbolic PDEs and then solving the above problems via the backstepping approach. In \cite{Kri:2009} the PDE model for the actuator and the sensor is a 1D diffusion equation, while in \cite{Kri:2009a} it is a 1D wave equation. In both these works, the output of the PDE is the boundary value of its state (Dirichlet measurement). The results in \cite{Kri:2009} and \cite{Kri:2009a} were extended in \cite{SuKr:2010} by studying interconnections in which the output of the PDE is the boundary value of the spatial derivative of its state (Neumann measurement). The ODE plants in \cite{Kri:2009}, \cite{Kri:2009a}, \cite{KrSm:2008} and \cite{SuKr:2010} have a single input and a single output. The paper \cite{LiKr:2010} considers plants with multiple inputs and outputs, with the actuator and sensor models being a set of 1D wave PDEs. The controllers that solve the stabilization problem in the above works are of the state feedback form. Recently, a dynamic output feedback controller was proposed in \cite{SaGaKr:2018} for solving the stabilization problem when the  PDE (actuator) is either a first-order hyperbolic equation or a 1D diffusion equation. In \cite{ZhGuWu:16}, combining the backstepping approach with the active disturbance rejection control method, an output feedback controller has been developed for stabilizing a wave PDE and ODE cascade system subject to boundary disturbance.

In this paper, we will solve the aforementioned stabilization problem for PDE-ODE (actuator-plant) cascade systems and the estimation problem for ODE-PDE (plant-sensor) cascade systems using the Sylvester equation. To explain our approach to solving the stabilization problem, let us suppose that the actuator model is also an ODE. Then the cascade system can be written as \vspace{-1mm}
\begin{equation} \label{plant-act}
 \dot w(t) = E w(t) + F C z(t), \qquad \dot z(t) = A z(t) + B u(t), \vspace{-1mm}
\end{equation}
where $w(t)\in\rline^n$ and $z(t)\in\rline^p$ are the states of the plant and the actuator, $u(t)\in\rline^m$ is the input and $C z(t)\in\rline^q$ is the actuator output and $E\in \rline^{n\times n}$, $A\in \rline^{p\times p}$, $B\in \rline^{p\times m}$, $C\in\rline^{q\times p}$ and $F\in\rline^{n\times q}$. Under the state transformation $\bbm{w & z} \to \bbm{p=w+\Pi z & z}$, where $\Pi\in\rline^{n\times p }$ is a solution to the Sylvester \vspace{-1mm} equation
$$ E \Pi = \Pi A +FC, \vspace{-1mm} $$
the state matrix of the cascade system \eqref{plant-act} becomes diagonal:
\vspace{-1mm}
$$ \bbm{\dot p(t) \\ \dot z(t)} = \bbm{E & 0 \\ 0 & A} \bbm{p(t) \\ z(t)} + \bbm{\Pi B \\ B}  u(t). \vspace{-1mm}$$
Suppose that $A$ is Hurwitz (i.e. the actuator model is stable) and the pair $(E,\Pi B)$ is stabilizable, so that $E+\Pi B K$ is Hurwitz for some $K\in\rline^{m\times n}$. Then the control law $u=K p$ stabilizes the above system, i.e. $u=Kw+K\Pi z$ is a stabilizing state feedback control law for the cascade system \eqref{plant-act}. In Section \ref{sec3}, we apply the above approach of diagonalizing the state matrix of the cascade system, to solve the stabilization problem for PDE actuator models belonging to the class of regular linear systems (RLSs). This approach reduces the stabilization problem to a problem of solving an appropriate Sylvester equation with unbounded operators and then stabilizing a finite-dimensional system, see Theorem \ref{th:act:stab}. In Section \ref{sec4}, we use an analogous approach to reduce the estimation problem for the ODE-PDE cascade system, when the PDE system is a RLS, to a problem of solving an appropriate Sylvester equation with unbounded operators and then constructing an estimator for a finite-dimensional system, see Theorem \ref{th:sen:det}.

Sylvester equations with unbounded operators play a central role in the state space approach to the output regulation of RLSs, see \cite{linreg}, \cite{Deu:11}, \cite{NaGiWe:14}, \cite{Pau:16}, \cite{XuDu:17} and references therein. This is due to the natural occurrence of ODE-PDE (exosystem-plant) cascade systems in the output regulation problem for RLSs. In fact, the Sylvester equation based diagonalization approach for stabilizing PDE-ODE cascade systems discussed in the previous paragraph was used in \cite[Theorem 13]{HaPo:10} to design observer-based controllers for solving the output regulation problem. In \cite{HaPo:10}, it is assumed that the control and observation operators of the PDE plant are bounded and the eigenvalues of the state matrix of the exosystem are on the imaginary axis. By relaxing the first assumption, the controller design technique and the associated diagonalization approach in \cite{HaPo:10} were generalized in \cite[Theorem 15]{Pau:16} by allowing the PDE plant to be any RLS with possibly unbounded control and observation operators. Furthermore, the Sylvester equation based diagonalization approach for building observers for ODE-PDE cascade systems, referred to as the `analogous approach' in the previous paragraph, is used implicitly in the controller design in \cite[Theorem 12]{Pau:16}. Recently, this `analogous approach' was used directly in \cite{XuDu:17} to construct observers for ODE-PDE (exosystem-plant) cascade systems, assuming that the control operator for the PDE plant is bounded and the eigenvalues of the state matrix of the exosystem are simple and lie on the imaginary axis. This work highlighted the advantage of the diagonalization approach by explicitly demonstrating how it simplifies the estimation problem for ODE-PDE cascade systems to an estimation problem for ODE systems, and thereby inspired the developments in the current work.

In this paper, we use the Sylvester equation based diagonalization approach to present a unified framework for constructing output feedback controllers for stabilizing PDE-ODE cascade systems in Section \ref{sec3} and observers for ODE-PDE cascade systems in Section \ref{sec4}. We let the PDE system be any stable (or easily stabilizable) RLS and the ODE system need not be marginally stable (unlike in the regulator theory). We derive necessary and sufficient conditions for verifying the solvability of the stabilization and estimation problems. We prove that the controller solving the stabilization problem is robust to certain unbounded perturbations of the PDE. The regularity assumption on the PDE system can be relaxed, see Remarks \ref{rm:act:nonreg} and \ref{rm:sen:nonreg}. Using these results we can solve the robust stabilization and estimation problems for (almost) all the PDE-ODE and ODE-PDE cascade systems which have been considered in the literature using the backstepping approach, see Section \ref{sec6} for a detailed discussion. In Section \ref{sec5}, we illustrate the results in Section \ref{sec3} using a 1D diffusion equation and the results in Section \ref{sec4} using a 1D wave equation. We remark that for 1D constant coefficient PDEs, it is straight forward to solve the Sylvester equation and construct the desired controllers and observers, see Remark \ref{rm:sylsol}.

The current paper is a significantly expanded version of the conference paper \cite{Nat:19}. In \cite{Nat:19} only the stabilization problem was considered, for which only a state feedback controller was developed under an assumption that is hard to verify. The proofs in \cite{Nat:19} were either shortened or omitted due to space constraints and the robustness of the controller was also not studied.

{\em Notation}: Define $\cline^-_\o=\{s\in\cline\big| \Re s<\o\}$ and $\cline^+_\o=\{s\in\cline\big| \Re s>\o\}.$ The closure of $\cline^-_\o$ and $\cline^+_\o$ in $\cline$ are denoted by $\overline{\cline^-_\o}$ and $\overline{\cline^+_\o}$. When $\o=0$, we drop the subscript. Let $X$ and $Y$ be Hilbert spaces. Then $\Lscr(X,Y)$, written as $\Lscr(X)$ if $X=Y$, denotes the space of bounded linear operators from $X$ to $Y$. The space of $X$-valued locally square integrable functions on $[0,\infty)$ is denoted as $L^2_{\rm loc}([0,\infty);X)$. For each $\alpha\in\rline$, the space $L^2_\alpha([0,\infty);X)$ $=\{ u\in L^2_{\rm loc}([0,\infty);X) \big| \int_0^\infty e^{-2\alpha t}\|u(t)\|^2\dd t <\infty\}$ is a Hilbert space with norm being the square root of the integral in the expression. For a linear operator $A:D(A)\subset X\to X$, where $D(A)$ is the domain of $A$, let $\sigma(A)$ be its spectrum and $\rho(A)$ its resolvent set. For a Banach space $X$, $H^\infty(X)$ is the Banach space of $X$-valued bounded analytic functions on $\cline^+$ with the sup norm.  Let $I_X$, or just $I$ when $X$ is clear, denote the identity operator on the space $X$.
\vspace{-5mm}

\section{Regular linear systems} \label{sec2}
\setcounter{equation}{0} 
\vspace{-1mm}

\ \ \ In this section, we summarize some results on regular linear systems and their feedback interconnections. For more details, see \cite{obs_book}, \cite{Wei:94a}, \cite{Wei:94b} and \cite{WeCu:97}.

Let $Z$, $U$ and $Y$ be Hilbert spaces. Let $A$ be the generator of a strongly continuous semigroup $\tline$ on $Z$ with growth bound $\o_\tline$. The semigroup $\tline$ (or equivalently $A$) is exponentially stable if $\o_\tline<0$. For some $\beta\in\rho(A)$, let $Z_1$ be the domain of $A$ with the norm $\|z\|_1=\|(\beta I - A)z\|$ and let $Z_{-1}$ be the completion of $Z$ with respect to the norm $\|z\|_{-1}=\|(\beta I - A)^{-1}z\|$. Let $B\in\Lscr(U,Z_{-1})$ be an admissible control operator for $\tline$. Let $C\in\Lscr(Z_1,Y)$ be an admissible observation operator for $\tline$ and let $C_\L$ be its $\L$-extension with respect to $A$. Then for each $\alpha> \o_\tline$ there exists $K_\alpha, M_\alpha \geq 0$ such that \vspace{-1mm}
\begin{equation} \label{eq:Best}
   \|(sI-A)^{-1}B\|_{\Lscr(U,Z)} \m\leq\m \frac{K_\alpha}{\sqrt{\Re s-\alpha}} \FORALL s\in\cline_\alpha^+, \vspace{-2mm}
\end{equation}
\begin{equation} \label{eq:Cest}
   \|C(sI-A)^{-1}\|_{\Lscr(Z,Y)} \m\leq\m \frac{M_\alpha}{\sqrt{\Re s-\alpha}} \FORALL s\in\cline_\alpha^+. \vspace{-1mm}
\end{equation}

Suppose that (i) $C_\L(sI-A)^{-1}B$ exists for each $s\in\rho(A)$ and (ii) $\sup_{s\in\cline_\alpha^+} \|C_\L(sI-A)^{-1} B\|_{\Lscr(U,Y)} <\infty$ for any $\alpha>\o_\tline$, then the triple $(A,B,C)$ is said to be regular. The regular linear system (RLS) $\Sigma$ corresponding to a regular triple $(A,B,C)$ and a feedthrough operator $D\in \Lscr(U,Y)$ is the pair of equations \vspace{-1mm}
\begin{align}
 \dot z(t) &= A z(t) + B u(t), \label{eq:RLSstate}\\[0.5ex]
 y(t) &= C_\L z(t) + D u(t). \label{eq:RLSoutput}\\[-4ex]\nonumber
\end{align}
The operators $(A,B,C,D)$ are the generating operators (GOs) of $\Sigma$, $A$ is the state operator and $Z$, $U$ and $Y$ are the state, input and output spaces, respectively. The RLS $\Sigma$ is exponentially stable if $A$ is exponentially stable. For each initial state $z(0)\in Z$ and input $u\in L^2_{\rm loc}([0,\infty);U)$, the state trajectory $z$ of $\Sigma$ (or \eqref{eq:RLSstate}) is \vspace{-1mm}
\begin{equation*}
 z(t) \m=\m \tline_t z(0) + \int_0^t \tline_{t-\tau}B u(\tau)\dd \tau \FORALL t\geq0. \vspace{-1mm}
\end{equation*}
This trajectory is the unique function in $C([0,\infty);Z) \cap H^1_{\rm loc} ([0,\infty); Z_{-1})$ which satisfies \eqref{eq:RLSstate} in $Z_{-1}$ for almost all $t\geq0$. Moreover, $z(t)\in D(C_\L)$ for almost all $t\geq0$ and \eqref{eq:RLSoutput} defines an output $y\in L^2_{\rm loc}([0,\infty);Y)$. The transfer function of $\Sigma$ is \vspace{-1mm}
\begin{equation} \label{eq:rlstf}
 \GGG(s) = C_\L(sI-A)^{-1}B+D \FORALL s\in\cline_{\o_\tline}^+. \vspace{-1mm}
\end{equation}
For each $\o>\o_\tline$, the map $\GGG:\cline_{\o}^+\to\Lscr(U,Y)$ is bounded. If $u\in L^2_\alpha([0,\infty);Y)$, then the output $y\in L^2_\gamma([0,\infty);Y)$ for each $\gamma>\max\{ \alpha,\o_\tline\}$ and $\hat y(s)=C(sI-A)^{-1}z(0)+\GGG(s)\hat u(s)$ for all $s\in\cline_{\gamma}^+$.


An operator $P\in\Lscr(Y,U)$ is an {\em admissible feedback operator} for the transfer function $\GGG$ in \eqref{eq:rlstf} if $[I_Y-P\GGG(s)]^{-1}$ exists and is bounded on $\cline_\alpha^+$ for some $\alpha\in\rline$.

\begin{definition} \label{def:stab}
The pair $(A,B)$ is {\em stabilizable} if there exists an admissible observation operator $K\in\Lscr(Z_1,U)$ for $\tline$ such that $(A,B,K)$ is a regular triple, $I\in\Lscr(U)$ is an admissible feedback operator for $K_\L(sI-A)^{-1}B$ and $A+BK_\L$ is the generator of an exponentially stable semigroup on $Z$.
\end{definition}

For any $K$ satisfying the conditions in the above definition, $u=K_\L z$ is called a stabilizing state feedback control law for \eqref{eq:RLSstate}. For each initial state $z(0)\in Z$, this control law defines an $u\in L^2([0,\infty);U)$ which ensures that the state trajectory $z$ of \eqref{eq:RLSstate} converges to zero. Suppose that $K\in\Lscr(Z,U)$, so that $K_\L=K$. Then, using \eqref{eq:Best}, it follows that $K$ satisfies all the conditions in the definition, except that the semigroup generated by $A+BK$ may not be exponentially stable. In particular, for some $\alpha\in\rline$ and each initial state $z(0)\in Z$ there exists a unique state trajectory $z\in L^2_\alpha([0,\infty);Z)$ of \eqref{eq:RLSstate} with $u=K z$. The operator $A+BK$ is exponential stable if this trajectory satisfies $\|z(t)\|\leq M e^{-\o t}\|z(0)\|$ for some $M,\o>0$ and each $t\geq0$.


\begin{definition} \label{def:det}
The pair $(C,A)$ is {\em detectable} if there exists an admissible control operator $L\in\Lscr(Y,Z_{-1})$ for $\tline$ such that $(A,L,C)$ is a regular triple, $I\in\Lscr(Y)$ is an admissible feedback operator for $C_\L(sI-A)^{-1}L$ and $A+LC_\L$ is the generator of an exponentially stable semigroup on $Z$.
\end{definition}
For $L$ as in the definition, since $(A,B,C)$ is a regular triple, the triple $(A+LC_\L,[B\ \ L],C)$ is regular. The state equation \vspace{-1mm}
\begin{equation} \label{eq:act:observer}
  \dot{\hat z}(t) = (A+LC_\L) \hat z(t) - L y(t) + (B+LD) u(t) \vspace{-1mm}
\end{equation}
is called an {\em observer} for \eqref{eq:RLSstate}-\eqref{eq:RLSoutput} and for every initial state $z(0)$ of \eqref{eq:RLSstate} and $\hat z(0)$ of \eqref{eq:act:observer} and $u\in L^2_{\rm loc}([0,\infty);U)$, we have $\|z(t)-\hat z(t)\|\leq Me^{-\o t} \|z(0)-\hat z(0)\|$.


For $k=1,2$, let $\Sigma_k$ be a RLS with state space $Z_k$, input space $U_k$, output space $Y_k$, input $u_k$, output $y_k$ and transfer function $\GGG_k$. Suppose that $Y_1=U_2$, $Y_2=U_1$, the identity operator $I_{U_1}$ is an admissible feedback operator for $\GGG_2\GGG_1$ and $I-D_2D_1$ is invertible. Then the feedback interconnection in Figure 1 is a RLS, denoted as $\Sigma_{fb}$, with state space $Z_1\times Z_2$, input space $U_1$, output space $Y_2$, input $v$ and output $y_2$. If the state operator of the RLS $\Sigma_{fb}$ is exponentially stable, then we call $\Sigma_2$ a {\em stabilizing output feedback controller} for $\Sigma_1$. \vspace{-4mm}

$$\hspace{40mm}\parbox{5in}{\includegraphics[scale=0.14]{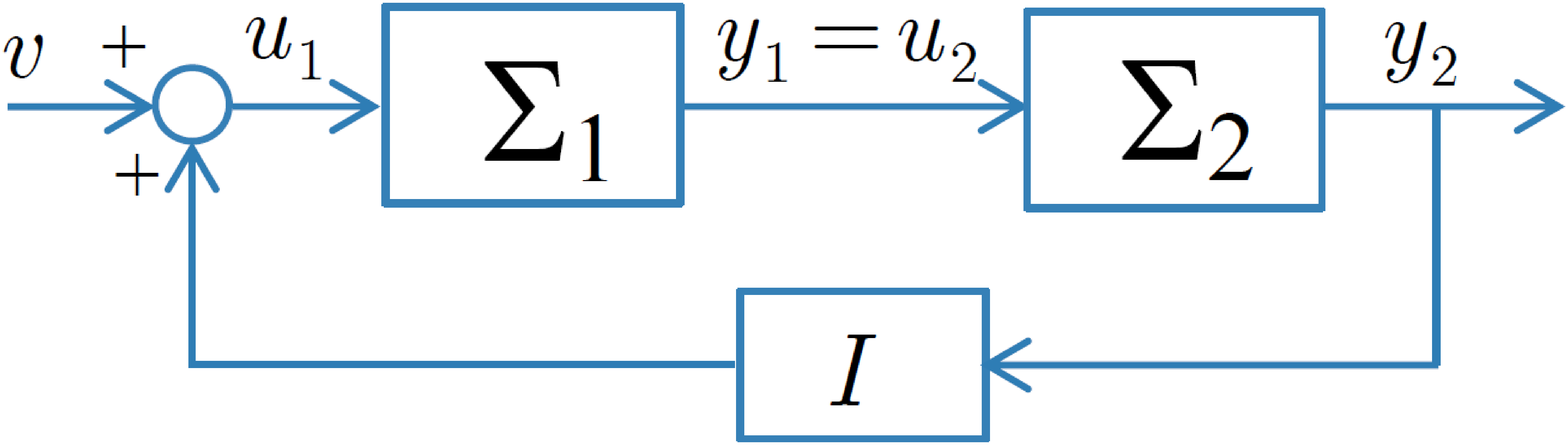}}$$
\centerline{ \parbox{5.2in}{Figure 1. Feedback interconnection of regular linear systems $\Sigma_1$ and $\Sigma_2$.\vspace{1mm}}}

\section{ODE plant with PDE actuator}
\label{sec3} \setcounter{equation}{0} 
\vspace{-1mm}

\ \ \ Consider a PDE-ODE cascade system in which the output of the PDE system drives the ODE system. The ODE models the plant dynamics, while the PDE models the actuator dynamics. The state dynamics of the cascade system is described by the following differential equations: for $t>0$ \vspace{-1mm}
\begin{align}
 \dot w(t) &= E w(t) + FC_\L z(t), \label{eq:act:ode}\\[0.5ex]
 \dot z(t) &= A z(t) + B u(t), \label{eq:act:pde}\\[-4ex]\nonumber
\end{align}
where $w(t)\in\rline^n$ is the plant state, $z(t)\in Z$ is the actuator state, $Z$ is a Hilbert space, $u(t)\in\rline^m$ is the input, $E\in \rline^{n\times n}$, $F\in\rline^{n\times q}$, $A$ is the generator of a strongly continuous semigroup $\tline$ on $Z$, $B\in \Lscr(\rline^m,Z_{-1})$ is an admissible control operator for $\tline$, $C\in\Lscr(Z_1,\rline^q)$ is an admissible observation operator for $\tline$ and $(A,B,C)$ is a regular triple. The admissibility of $B$ is not essential and can be relaxed, see Remark \ref{rm:act:nonreg}.
The output $y$ of the plant takes values in $\rline^p$ and is given by \vspace{-1mm}
\begin{equation} \label{eq:act:output}
 y(t) = G w(t) + H C_\L z(t),\qquad t\geq0, \vspace{-1mm}
\end{equation}
where $G\in\rline^{p\times n}$ and $H\in\rline^{p\times q}$.
For the PDE system (actuator), the output is $C_\L z$ and transfer function is \vspace{-2mm}
\begin{equation} \label{eq:act:RLStf}
 \GGG(s) = C_\L(sI-A)^{-1}B \FORALL s\in\cline^+_{\o_\tline}. \vspace{-2mm}
\end{equation}
The combined state space for the plant and actuator is $Z_{cs}=\rline^n\times Z$ and the state, control and observation operators for the combined dynamics (with input $u$, state $[w\ \ z]^\top$ and output $y$) are \vspace{-1mm}
\begin{equation} \label{eq:act:cas}
 A_{cs} = \bbm{E & FC_\L \\ 0 & A}, \qquad B_{cs} = \bbm{0\\ B}, \qquad C_{cs} = \bbm{G & HC_\L}. \vspace{-1mm}
\end{equation}
From the feedback theory for regular linear systems \cite[Lemma 5.1]{WeCu:97} it follows that the cascade system \eqref{eq:act:ode}-\eqref{eq:act:output} is a RLS, denoted as $\Sigma_{cs}$, with generating operators  $(A_{cs}, B_{cs}, C_{cs},0)$, input space $\rline^m$, state space $Z_{cs}$ and output space $\rline^p$.


We suppose, with no loss of generality, that $E$ is of the form \vspace{-2mm}
\begin{equation} \label{eq:act:Edecom}
 E = \bbm{E_1 & 0\\ 0 & E_2}, \vspace{-2mm}
\end{equation}
where $E_1\in\rline^{n_1\times n_1}$, $\sigma(E_1) \subset \overline{\cline^+}$, $E_2\in\rline^{n_2\times n_2}$ and $\sigma(E_2) \subset {\cline^-}$. The corresponding partitioning of $w$, $F$ and $G$ are \vspace{-2mm}
\begin{equation} \label{eq:act:part}
 w=\bbm{w_1\\ w_2},\qquad F = \bbm{F_1 \\ F_2}, \qquad G=\bbm{G_1 & G_2}. \vspace{-2mm}
\end{equation}
We will derive a stabilizing state feedback control law $u=K_{cs}[w\ \ z]^\top$ for \eqref{eq:act:ode}-\eqref{eq:act:pde} in Theorem \ref{th:act:stab}. A stabilizing output feedback controller for $\Sigma_{cs}$ is presented in Theorem \ref{th:act:det}. These results can be extended to derive stabilizing controllers for the system \eqref{eq:act:ode}-\eqref{eq:act:pde} modified to include a term $J u$ in \eqref{eq:act:ode}, see Remark \ref{rm:act:notcas}. We will need the following two assumptions.

\begin{assumption} \label{as:act:exp}
The semigroup $\tline$ (or equivalently $A$) is exponentially stable. \vspace{-1mm}
\end{assumption}

Assumption \ref{as:act:exp} is made to simplify the presentation and it is no more restrictive than the requirement that the pair $(A,B)$ be stabilizable. Indeed, if $A$ is not stable and $(A,B)$ is stabilizable, consider the cascade interconnection of \eqref{eq:act:ode} and  \vspace{-1mm}
\begin{equation} \label{eq:act:repde}
 \dot z(t) = (A+BK_\L) z(t) + B u(t), \vspace{-1mm}
\end{equation}
where $K$ is as in Definition \ref{def:stab}. A state feedback control law $K_1w+K_{2\L}z+K_\L z$ is stabilizing for \eqref{eq:act:ode}-\eqref{eq:act:pde} if and only if $K_1 w+K_{2\L}z$ is a stabilizing state feedback control law for \eqref{eq:act:ode}, \eqref{eq:act:repde}. Hence when $A$ is not stable, we can work with \eqref{eq:act:ode}, \eqref{eq:act:repde} for which Assumption \ref{as:act:exp} holds, instead of \eqref{eq:act:ode}-\eqref{eq:act:pde}. We remark that when $B$ is bounded, Assumption \ref{as:act:exp} is no more conservative than the natural assumption that the system \eqref{eq:act:ode}-\eqref{eq:act:pde} is stabilizable. This follows from the observation that the stabilizability of \eqref{eq:act:ode}-\eqref{eq:act:pde} implies the optimizability of $(A,B)$, which then implies the stabilizability of $(A,B)$ \cite{Cur-Zw} (for unbounded $B$ the latter implication is not known \cite{WeRe:00}). In the particular case in which the unstable subspace of $A$ is finite-dimensional, we can combine it with the unstable subspace of $E$, redefine $A$, $B$, $C$, $E$ and $F$ suitably and then work with \eqref{eq:act:ode}-\eqref{eq:act:pde} (with redefined operators) for which Assumption \ref{as:act:exp} holds, see Example 5.1 for an illustration of this approach.

\begin{assumption} \label{as:act:stab}
$v^\top F_1\GGG(\l)\neq0$ for each eigenvalue $\l\in\sigma(E_1)$ and nonzero vector $v\in\rline^{n_1}$ satisfying $v^\top E_1=\l v^\top$. \vspace{-1mm}
\end{assumption}

Note that $\GGG(\l)$ exists for all $\l\in\sigma(E_1)$ since $\sigma(E_1) \subset \cline^+_{ \o_\tline}\subset\rho(A)$ by Assumption \ref{as:act:exp}. Assumption \ref{as:act:stab} implies that $v^\top F_1\neq0$ for each left eigenvector $v^\top$ of $E_1$, which in turn implies via the Hautus test that the pair $(E_1,F_1)$ is stabilizable. In Proposition \ref{pr:act:stabcon} we will show that when $A$ is exponentially stable, the pair $(A_{cs},B_{cs})$ is stabilizable if and only if Assumption \ref{as:act:stab} holds. So when $A$ is not exponentially stable, in light of the discussion below Assumption \ref{as:act:exp}, the pair $(A_{cs},B_{cs})$ is stabilizable if (and also {\em only if} when $B$ is bounded) $(A,B)$ is stabilizable and for some $K$ as in Definition \ref{def:stab}, Assumption \ref{as:act:stab} holds with $\GGG_K(\l)=C_\L(\l I-A-BK_\L)^{-1}B$ in place of $\GGG(\l)$. If $\GGG(\l)$ exists for a $\l\in\sigma(E_1)$, then it is easy to check that
$ \GGG_K(\l)=\GGG(\l)(I+K_\L(\l I-A-BK_\L)^{-1}B)$ and $\GGG(\l)=\GGG_K(\l)(I-K_\L(\l I-A)^{-1}B), $
which implies that $v^\top F_1\GGG(\l)\neq0$ if and only if $v^\top F_1\GGG_K(\l)\neq0$. Therefore, if $\GGG(\l)$ exists for each $\l\in\sigma(E_1)$, the pair $(A_{cs},B_{cs})$ is stabilizable if $(A,B)$ is stabilizable and Assumption \ref{as:act:stab} holds, see Example 5.1 for an illustration.

Next we present a result on the existence of solutions to Sylvester equations with unbounded operators. This result has been established in \cite{Pau:16} assuming that $\sigma(\Escr)$ lies on the imaginary axis.

\begin{framed} \vspace{-3mm}
\begin{lemma} \label{pr:act:syl}
Let $\Ascr$ be the generator of an exponentially stable strongly continuous semigroup $\sline$ on a Hilbert space $X$. Let $\Escr\in\rline^{n\times n}$ be such that $\sigma(\Escr)\subset \overline{\cline^+}$. Let $Q\in \Lscr(X_1,\rline^n)$ be an admissible observation operator for $\sline$. Then there exists a linear map $\Pi:\Ascr D(Q_\L)\to\rline^n$ with $\Pi\in \Lscr(X,\rline^n)$ such that \vspace{-2mm}
\begin{equation} \label{eq:act:syl}
 \Escr\Pi x = \Pi \Ascr x + Q_\L x \FORALL x\in D(Q_\L). \vspace{-3mm}
\end{equation}

Furthermore, if $P\in\Lscr(\rline^m,X_{-1})$ is an admissible control operator for $\sline$ and $(\Ascr,P,Q)$ is a regular triple, then $\Pi P\in \Lscr(\rline^m,\rline^n)$. \vspace{-3mm}
\end{lemma}
\end{framed}
\vspace{-3mm}
\begin{proof}
Observe that $e^{-\Escr t}$ can be written as follows: \vspace{-1mm}
\begin{equation} \label{eq:act:matexp}
 e^{-\Escr t} = \sum_{k=1}^{v}\sum_{j=0}^{r} E_{kj} e^{-\l_k t} \frac{t^j}{j!}, \vspace{-1mm}
\end{equation}
where each $E_{kj}\in\rline^{n\times n}$ is a constant matrix and $\l_k\in\overline{\cline^+}$ is an eigenvalue of $\Escr$. Taking the derivative of \eqref{eq:act:matexp} with respect to $t$ gives \vspace{-1mm}
$$ -\Escr \sum_{k=1}^{v}\sum_{j=0}^{r} E_{kj} e^{-\l_k t} \frac{t^j}{j!} = \sum_{k=1}^{v}\sum_{j=0}^{r}\left(-E_{kj}\l_k  + E_{k\, j+1} \right) e^{-\l_k t} \frac{t^j}{j!}, \vspace{-1mm}$$
where $E_{k\, r+1}=0$ by definition. Comparing the coefficients of $e^{-\l_k t}t^j$ on both sides it then follows that for $k\in\{1,2,\ldots v\}$ and $j\in\{0,1,\ldots r\}$, \vspace{-1mm}
\begin{equation} \label{eq:act:matexpid}
 \Escr E_{kj} = \l_k E_{kj} - E_{k\,j+1}. \vspace{-1mm}
\end{equation}
Define $\Pi\in\Lscr(X,\rline^n)$ as follows: \vspace{-1mm}
\begin{equation} \label{eq:act:Pisum}
 \Pi = \sum_{k=1}^v\sum_{j=0}^r E_{kj} Q_\L (\l_k-\Ascr)^{-1-j}. \vspace{-1mm}
\end{equation}
Then $\Pi$ maps $\Ascr D(Q_\L)$ to $\rline^n$ and solves \eqref{eq:act:syl}. Indeed, for any $x\in D(Q_\L)$, \vspace{-1mm}
\begin{align*}
 \Pi\Ascr x &= \sum_{k=1}^v\sum_{j=0}^r \l_k E_{kj} Q_\L (\l_k-\Ascr)^{-1-j}x - E_{kj} Q_\L (\l_k-\Ascr)^{-j}x  \\
 &= -\sum_{k=1}^v E_{k0}Q_\L x + \sum_{k=1}^v\sum_{j=0}^r (\l_k E_{kj}-E_{k\, j+1}) Q_\L (\l_k-\Ascr)^{-1-j}x .\\[-4ex]
\end{align*}
Using $ \sum_{k=1}^{v} E_{k0} = I$, which follows by letting $t=0$ in \eqref{eq:act:matexp}, and \eqref{eq:act:matexpid} it follows that the expression on the last line is $\Escr\Pi x - Q_\L x$.

Finally, if $(\Ascr,P,Q)$ is a regular triple, then by definition $Q_\L (sI-\Ascr)^{-1}P\in\Lscr(\rline^m,\rline^n)$ for each $s\in\rho(\Ascr)$. This, the fact that $\rho(A)\cap\sigma(\Escr)= \emptyset$ and the expression for $\Pi$ in \eqref{eq:act:Pisum} imply that $\Pi P\in\Lscr(\rline^m,\rline^n)$.
\end{proof}

Next we present a stabilizing state feedback control law for the PDE-ODE system \eqref{eq:act:ode}-\eqref{eq:act:pde}. Recall the notation $E_1$, $E_2$, $F_1$, $F_2$, $w_1$ and $w_2$ from \eqref{eq:act:Edecom} and \eqref{eq:act:part}. \vspace{-1mm}

\begin{framed} \vspace{-3mm}
\begin{theorem} \label{th:act:stab}
Consider the PDE-ODE cascade system \eqref{eq:act:ode}-\eqref{eq:act:pde}. Suppose that Assumption \ref{as:act:exp} holds. Define \vspace{-2mm}
$$ A_1 = \bbm{E_2 & F_2 C_\L\\ 0 & A}, \qquad B_1 = \bbm{\ 0_{n_2\times m}\\ \!\!\!\!\!\!\!\!\!B}, \qquad C_1 = \bbm{0_{q\times n_2} & C}. \vspace{-2mm}$$
Then $A_1$ is the generator of an exponentially stable strongly continuous semigroup $\sline$ on $X=\rline^{n_2}\times Z$, the control operator $B_1\in\Lscr (\rline^m,X_{-1})$ and the observation operator $C_1\in\Lscr(X_1,\rline^q)$ are admissible for $\sline$ and the triple $(A_1,B_1,C_1)$ is regular. There exists $\Pi:A_1 D(C_{1\L})\to\rline^{n_1}$ with $\Pi\in \Lscr(X, \rline^{n_1})$ such that \vspace{-2mm}
\begin{equation} \label{eq:act:sylth}
 E_1 \Pi x = \Pi A_1 x + F_1 C_{1\L} x \FORALL x\in D(C_{1\L}) \vspace{-2mm}
\end{equation}
and $\Pi B_1\in\Lscr(\rline^m,\rline^{n_1})$.

Suppose that Assumption \ref{as:act:stab} also holds. Then the pair $(E_1,\Pi B_1)$ is stabilizable. Let $K\in \rline^{m\times n_1}$ be such that $E_1+\Pi B_1 K$ is Hurwitz. Then $u = Kw_1+K\Pi [w_2\ \ z]^\top$ is a stabilizing state feedback control law for \eqref{eq:act:ode}-\eqref{eq:act:pde}. Moreover, for all $\delta\in\rline$ sufficiently small, this control law also stabilizes the perturbed RLS \vspace{-4mm}
\begin{align}
 \dot w(t) &= E w(t) + FC_\L z(t), \label{eq:act:Rode}\\[0.5ex]
 \dot z(t) &= (A+\delta A) z(t) + B u(t). \label{eq:act:Rpde}\\[-5ex]\nonumber
\end{align}
\end{theorem}\vspace{-3mm}
\end{framed}
\vspace{-4mm}

\begin{proof}
The semigroup generated by $A_1$ on $X$ is $\sline_t=\sbm{e^{E_2 t} & \star\\ 0 & \tline_t}$ for all $t\geq0$, where the $\star$ denotes some non-zero entry. Since $\sigma(E_2) \subset\cline^-$ and $\tline$ is exponentially stable, $\sline$ is exponentially stable. All this and the admissibility of $B_1$ and $C_1$ and the regularity of the triple $(A_1,B_1,C_1)$ follow from the feedback theory for RLSs \cite[Lemma 5.1]{WeCu:97}. Since $(A_1,B_1,C_1)$ is regular and $F_1$ is a bounded map, we can conclude that $F_1C_1$ is an admissible observation operator for $\sline$, its $\L$-extension is $F_1C_{1\L}$ with $D(F_1C_{1\L})=D(C_{1\L})$ and $(A_1,B_1,F_1C_1)$ is regular. Hence applying Lemma \ref{pr:act:syl} with $\Escr=E_1$, $\Ascr=A_1$, $Q=F_1C_1$ and $P=B_1$, we get that there exists $\Pi\in\Lscr(X, \rline^{n_1})$ which solves \eqref{eq:act:sylth} and $\Pi B_1\in\Lscr(\rline^m, \rline^{n_1})$. From \eqref{eq:act:Pisum} we have $\Pi = \sum_{k=1}^v\sum_{j=0}^r E_{kj} F_1C_{1\L} (\l_k-A_1)^{-1-j}$ for some matrices $E_{kj}$ and $\l_k\in\sigma(E_1)$.

Suppose that Assumption \ref{as:act:stab} holds. Then the pair $(E_1,\Pi B_1)$ is stabilizable. Indeed, if not, then via the Hautus test there exists a $\l\in\sigma(E_1)$ and non-zero $v\in\rline^{n_1}$ such that \vspace{-1mm}
\begin{equation} \label{eq:act:Hau}
 v^\top E_1 = \l v^\top, \qquad v^\top \Pi B_1=0. \vspace{-1mm}
\end{equation}
Since $(A_1,B_1,C_1)$ is a regular triple, $(\l I - A_1)^{-1}B_1 U \subset D(C_{1\L})$. Choosing $x=(\l I-A_1)^{-1} B_1 u_1$ in \eqref{eq:act:sylth} with $u_1\in\rline^m$ and then applying $v^\top$ from the left to both sides of the resulting expression, we get using $C_{1\L}(\l I-A_1)^{-1}B_1=\GGG(\l)$ and the first equation in \eqref{eq:act:Hau} that \vspace{-2mm}
\begin{equation} \label{eq:act:vPiB1}
  v^\top\Pi B_1 u_1 = v^\top F_1 \GGG(\l)u_1 \FORALL u_1\in \rline^m.
  \vspace{-1mm}
\end{equation}
Using the second equation in \eqref{eq:act:Hau} it follows from \eqref{eq:act:vPiB1} that $v^\top F_1\GGG(\l)=0$, which contradicts Assumption \ref{as:act:stab}. Hence the pair $(E_1,\Pi B_1)$ is stabilizable.

Fix $K\in\Lscr(\rline^{n_1},\rline^m)$ such that $E_1+\Pi B_1 K$ is Hurwitz. Define $K_{cs}\in\Lscr(Z_{cs},\rline^m)$ by $K_{cs}[w\ \ z]^\top=Kw_1+K\Pi z_1$, where $z_1=[w_2\ \ z]^\top$. Recall that \eqref{eq:act:ode}-\eqref{eq:act:pde} can be written as $\dot \nu = A_{cs} \nu + B_{cs} u$, where $\nu = [w \ \ z]^\top$. Since $K_{cs}$ is bounded, it follows from the discussion below Definition \ref{def:stab} that for some $\alpha\in\rline$ and each initial state $[w(0)\ \ z(0)]^\top\in Z_{cs}$ there exists a unique state trajectory $[w \ \ z]^\top\in L^2_\alpha([0,\infty);Z_{cs})$ of \eqref{eq:act:ode}-\eqref{eq:act:pde} with $u=K_{cs}[w\ \ z]^\top$. Since $(A,B,C)$ is a regular triple, we have $C_\L z \in L^2_\gamma([0,\infty); \rline^q)$ for some $\gamma>\alpha$. Along this state trajectory, $w_1$ and $z_1$ satisfy \vspace{-1mm}
$$ \dot w_1(t) = E_1 w_1(t) + F_1 C_{1\L} z_1(t),\qquad
  \dot z_1(t) = A_1 z_1(t) + B_1 (Kw_1(t)+K\Pi z_1(t)) \vspace{-1mm}$$
in $\rline^{n_1}\times X_{-1}$, for almost all $t\geq0$ . Note that  $C_{1\L}=\bbm{0&C_\L}$ and hence $C_{1\L} z_1=C_\L z\in  L^2_\gamma([0,\infty); \rline^q)$.  Define $p_1=w_1+\Pi z_1$. Taking the Laplace transform of the above equations we get that for all $s\in\cline^+_{\max\{0,\gamma\}}\cap\rho(E_1)$ \vspace{-1mm}
\begin{align}
 \hat z_1(s) &= (sI-A_1)^{-1}z_1(0)+(sI-A_1)^{-1}B_1K\hat p_1(s), \nonumber\\[0.5ex]
 \hat p_1(s) &= (sI-E_1)^{-1}\big[\m F_1C_{1\L}+ (sI-E_1)\Pi\m\big] (sI-A_1)^{-1}\big[\m B_1 K\hat p_1(s) + z_1(0)\m\big]  \nonumber\\
 &\qquad + (sI-E_1)^{-1} w_1(0). \label{eq:act:p1hat}
\end{align}
Here {\em hat} denotes the Laplace transform. From \eqref{eq:act:sylth}, we have $F_1C_{1\L}+ (sI-E_1)\Pi=\Pi(sI-A_1)$. Using this in \eqref{eq:act:p1hat} we get \vspace{-1mm}
$$ \hat p_1(s) = (sI-E_1)^{-1}\Pi B_1K\hat p_1(s) +(sI-E_1)^{-1} p_1(0). \vspace{-1mm} $$
Hence $p_1$ satisfies the ODE \vspace{-2mm}
\begin{equation} \label{eq:act:peqn}
 \dot p_1(t) = (E_1+\Pi B_1K) p_1(t). \vspace{-1mm}
\end{equation}
The above equation can also be derived by proving that $\dd (\Pi z_1(t)) /\dd t=\Pi(\dd  z_1(t)/\dd t)$. Hence along the trajectory $[w \ \ z]^\top$, the transformed state $[p_1\ \ z_1]^\top$ satisfies \vspace{-1mm}
\begin{equation} \label{eq:act:pz}
 \bbm{\dot p_1(t)\\ \dot z_1(t)} = \bbm{E_1+\Pi B_1 K & 0\\  B_1 K & A_1}\bbm{p_1(t)\\z_1(t)}, \vspace{-1mm}
\end{equation}
with $p_1(0)=w_1(0)+\Pi w_1(0)$ and $z_1(0)=[w_2(0)\ \ z(0)]^\top$. Since $E_1+\Pi B_1 K$ and $A_1$ are both exponentially stable, it follows from the feedback theory of RLSs \cite[Lemma 5.1]{WeCu:97} that $\sbm{E_1+\Pi B_1 K & 0\\  B_1 K & A_1}$ is the generator of an exponentially stable strongly continuous semigroup on $\rline^{n_1}\times X$. Hence there exist $M_1,\o>0$ such that \vspace{-1mm}
\begin{equation} \label{eq:act:pzdec}
 \|p_1(t)\|+\|z_1(t)\| \leq M_1 e^{-\o t}(\|p_1(0)\|+\|z_1(0)\|) \FORALL t\geq0, \vspace{-1mm}
\end{equation}
which implies that there exists $M,\o>0$ such that \vspace{-1mm}
\begin{equation} \label{eq:act:wzdec}
  \|w(t)\|+\|z(t)\| \leq M e^{-\o t}(\|w(0)\|+\|z(0)\|) \FORALL t\geq0. \vspace{-1mm}
\end{equation}
It now follows from the discussion below Definition \ref{def:stab} that $K_{cs}[w\ \ z]^\top$ is a stabilizing state feedback control law for \eqref{eq:act:ode}-\eqref{eq:act:pde}, i.e. $A_{cs}+B_{cs}K_{cs}$ is exponentially stable.

We will now establish the robustness claim in the theorem. For each $\delta\in(-1,\infty)$ define $A_1^\delta$ and $A_{cs}^\delta$ similarly to $A_1$ and $A_{cs}$, but with $A+\delta A$ in place of $A$. Then $A_1^\delta$ is the generator of an exponentially stable semigroup $\sline^\delta$ given by $\sline^\delta_t =\sline_{(1+\delta)t}$ for all $t\geq0$. For any $\l\in\cline^+$ and integer $k\geq1$, the triangular structure of $A_1$ and $C_{1\L}=\bbm{0&C_\L}$ imply that $C_{1\L}(\l-A_1)^{-k}=[\ 0 \ \ C_{\L}(\l-A)^{-k}\ ]$. Using this and the expression for $\Pi$ we get \vspace{-3mm}
\begin{align}
 \Pi (A_1^\delta-A_1) &= \sum_{k=1}^v\sum_{j=0}^r E_{kj} C_{1\L} (\l_k-A_1)^{-1-j} (A_1^\delta-A_1)\nonumber\\
  &= \delta \sum_{k=1}^v\sum_{j=0}^r E_{kj} \bbm{0 & C_\L (\l_k-A)^{-1-j}A}. \label{eq:act:piadm}\\[-5ex]\nonumber
\end{align}
The admissibility of $[0 \ \ C_\L]$ for $\sline$ and \eqref{eq:act:piadm} imply that $C_2=\Pi (A_1^\delta-A_1)/\delta$ is an admissible observation for $\sline^\delta$ and $C_{2\L}=C_2$. The regularity of the triple $(A_1^\delta,B_1,C_2)$ follows from the regularity of the triple $(A_1,B_1,C_1)$. The system \eqref{eq:act:Rode}-\eqref{eq:act:Rpde} can be written as $\dot \nu = A_{cs}^\delta \nu + B_{cs} u$, where $\nu = [w \ \ z]^\top$, and $B_{cs}$ is admissible for the semigroup generated by $A_{cs}^\delta$ \cite[Lemma 5.1]{WeCu:97}. Since $K_{cs}$ is bounded, it follows from the discussion below Definition \ref{def:stab} that for each initial state $[w(0)\ \ z(0)]^\top\in Z_{cs}$ there exists a unique state trajectory $[w \ \ z]^\top$ for \eqref{eq:act:Rode}-\eqref{eq:act:Rpde} with input $u=K_{cs}[w\ \ z]^\top$. By  adapting the arguments used to derive \eqref{eq:act:pz}, we get that along this state trajectory the transformed state $[p_1=w_1+\Pi z_1\ \ z_1]^\top$ satisfies \vspace{-1mm}
\begin{equation} \label{eq:act:p1z1}
 \bbm{\dot p_1(t)\\ \dot z_1(t)} = \bbm{E_1+\Pi B_1 K & \Pi (A_1^\delta-A_1)\\  B_1 K & A_1^\delta}\bbm{p_1(t)\\z_1(t)}, \vspace{-1mm}
\end{equation}
with $p_1(0)=w_1(0)+\Pi w_1(0)$ and $z_1(0)=[w_2(0)\ \ z(0)]^\top$.

Consider the RLS $\Sigma_1^\delta$ with GOs $(A_1^\delta,B_1,C_2,0)$ and transfer function $\GGG_1^\delta$ and the RLS $\Sigma_2^\delta$ with GOs $(E_1+\Pi B_1 K, \delta I_{\rline^{n_1}}, K, 0)$ and transfer function $\GGG_2^\delta$. Since $\Sigma_1^\delta$ and $\Sigma_2^\delta$ are exponentially stable, their positive feedback interconnection $\Sigma_{fb}^\delta$ is also an exponentially stable RLS if $(I-\GGG_1^\delta\GGG_2^\delta)^{-1}\in H^\infty (\Lscr(\rline^{n_1}))$ \cite[Proposition 4.6]{WeCu:97}. The exponential stability of $\Sigma_1^\delta$ implies that $\GGG_1^0\in H^\infty (\Lscr(\rline^m,\rline^{n_1}))$ and we have $\GGG_1^\delta(s)=(1+\delta)^{-1}\GGG_1^0(s(1+\delta)^{-1})$. Therefore, for $\delta$ belonging to any compact subset of $(-1,\infty)$,
$\|\GGG_1^\delta\|_{ H^\infty(\Lscr(\rline^m,\rline^{n_1}))}$ can be bounded by a constant independent of $\delta$. In addition, $\lim_{\delta \to 0}\|\GGG_2^\delta\|_{H^\infty(\Lscr(\rline^{n_1},\rline^m))}=0$. Therefore $(I-\GGG_1^\delta\GGG_2^\delta )^{-1}\in H^\infty (\Lscr(\rline^{n_1}))$ for all $\delta$ sufficiently small. Consequently $\sbm{E_1+\Pi B_1 K & \Pi (A_1^\delta-A_1)\\  B_1 K & A_1^\delta}$, being the state operator of $\Sigma_{fb}^\delta$, is exponentially stable. It now follows from \eqref{eq:act:p1z1} that $[p_1\ \ z_1]^\top$ satisfies an estimate of the form \eqref{eq:act:pzdec} and so $[w\ \ z]^\top$ satisfies an estimate of the form \eqref{eq:act:wzdec}. Hence, according to the discussion below Definition \ref{def:stab}, for $\delta$ small $K_{cs}[w\ \ z]^\top$ is a stabilizing state feedback control law for \eqref{eq:act:Rode}-\eqref{eq:act:Rpde}, i.e. $A_{cs}^\delta+ B_{cs}K_{cs}$ is exponentially stable. \vspace{-1mm}
\end{proof}

Theorem \ref{th:act:stab} shows that Assumption \ref{as:act:stab} is sufficient for the existence of a stabilizing control law for the PDE-ODE system \eqref{eq:act:ode}-\eqref{eq:act:pde}. The next proposition establishes that this assumption is also necessary. \vspace{-1mm}

\begin{framed} \vspace{-3mm}
\begin{proposition} \label{pr:act:stabcon}
Consider the PDE-ODE system \eqref{eq:act:ode}-\eqref{eq:act:pde}.\!
Let Assumption \ref{as:act:exp} hold. Then the pair $(A_{cs}, B_{cs})$ is stabilizable if and only if Assumption \ref{as:act:stab} holds.
\end{proposition}
\vspace{-3mm}
\end{framed}
\vspace{-7mm}

\begin{proof}
Suppose that Assumption \ref{as:act:stab} holds. We have shown in Theorem \ref{th:act:stab} that $(A_{cs}, B_{cs})$ is stabilizable and found  $K_{cs}$ such that $A_{cs}+B_{cs}K_{cs}$ is exponentially stable.

Conversely, suppose that the pair $(A_{cs},B_{cs})$ is stabilizable. If Assumption \ref{as:act:stab} does not hold, then there exists a non-zero  $v\in\rline^{n_1}$ such that $v^\top E_1=\l v^\top$ for some $\l\in\sigma(E_1)$ and $v^\top F_1 \GGG(\l)=0$. It now follows from \eqref{eq:act:vPiB1} that $v^\top\Pi B_1=0$ which, via the Hautus test, implies that the pair $(E_1,\Pi B_1)$ is not stabilizable. Consequently there exists a $p_0\in \rline^{n_1}$ such that the state trajectory of \vspace{-2mm}
\begin{equation} \label{eq:act:p1contr}
 \dot p_1(t) = E_1 p_1(t) + \Pi B_1 u(t), \qquad p_1(0)=p_0, \vspace{-2mm}
\end{equation}
satisfies \vspace{-2mm}
\begin{equation} \label{eq:act:limp1}
 \liminf_{t\to\infty}\|p_1(t)\|>0 \FORALL u\in L^2([0,\infty); \rline^m). \vspace{-1mm}
\end{equation}
On the other hand, since the pair $(A_{cs},B_{cs})$ is stabilizable, there exists a $\tilde u\in L^2([0,\infty); \rline^m)$ such that the state trajectory $[w\ \ z]^\top$ of \eqref{eq:act:ode}-\eqref{eq:act:pde} for the input $u=\tilde u$ and initial state $w(0)=[w_1(0)\ \ w_2(0)]^\top=[p_0\ \ 0]^\top$ and $z(0)=0$ satisfies $\lim_{t\to\infty} (\|w(t)\|+\|z(t)\|) = 0$, see comment below Definition \ref{def:stab}. Via arguments similar to those used to derive \eqref{eq:act:peqn}, it can be shown that along this trajectory $p_1$ defined as  $w_1+\Pi\,[w_2 \ \ z]^\top$ solves \eqref{eq:act:p1contr} with $u=\tilde u$. Clearly $\lim_{t\to\infty} \|p_1(t)\| = 0$ (as $w(t), z(t)$ decay to 0), which contradicts \eqref{eq:act:limp1}. So Assumption \ref{as:act:stab} must hold. \vspace{-1mm}
\end{proof}

The next theorem presents an observer-based stabilizing output feedback controller $\Sigma_c$ for the PDE-ODE cascade system \eqref{eq:act:ode}-\eqref{eq:act:output}. Recall that this system is a RLS, denoted as $\Sigma_{cs}$, with GOs $(A_{cs},B_{cs},C_{cs},0)$ introduced in \eqref{eq:act:cas}.

\begin{assumption} \label{as:act:det}
The pair $(G,E)$ is detectable.
\end{assumption}

From \eqref{eq:act:Edecom} and \eqref{eq:act:part} it follows that Assumption \ref{as:act:det} is equivalent to the detectability of the pair $(G_1,E_1)$ and if $E_1+L_1G_1$ is Hurwitz, then so is $E+L G$, where $L=[L_1 \ \ 0]^\top$. Recall the control law $u=K w_1+K\Pi z_1$ proposed in Theorem \ref{th:act:stab} which can be written as $u = K_1 w+K_2 z$ with $K_1\in\Lscr(\rline^n, \rline^m)$ and \vspace{-1mm} $K_2\in\Lscr(Z,\rline^m)$.

\begin{framed} \vspace{-3mm}
\begin{theorem} \label{th:act:det}
Consider the PDE-ODE cascade system \eqref{eq:act:ode}-\eqref{eq:act:output}. Suppose that Assumptions  \ref{as:act:exp}, \ref{as:act:stab} and \ref{as:act:det} hold. Let $L_1\in\Lscr(\rline^p,\rline^{n_1})$ be such that $E_1+L_1G_1$ is Hurwitz. Define $L=[L_1 \ \ 0]^\top\in\Lscr(\rline^p,\rline^n)$. Let $u= K_1 w+K_2 z$ be the stabilizing state feedback control law for \eqref{eq:act:ode}-\eqref{eq:act:pde} proposed in Theorem \ref{th:act:stab}. Then the quadruple of operators $(A_c,B_c,C_c,D_c)$ defined as \vspace{-1mm}
$$  A_c = \bbm{E+LG & (F+LH)C_\L \\ BK_1 & A+B K_2}, \quad B_c = \bbm{ -L\\0}, \quad C_c = \bbm{K_1 & K_2}, \quad D_c =0, \vspace{-1mm}$$
are the GOs of a RLS $\Sigma_c$ with input space $\rline^p$, state space $Z_{cs}$ and output space $\rline^m$ and $\Sigma_c$ is a stabilizing output feedback controller for $\Sigma_{cs}$.

For each $\delta\in(-1,\infty)$, let $\Sigma_{cs}^\delta$ be the RLS with GOs $(A_{cs}^\delta, B_{cs}, C_{cs}, 0)$, where $A_{cs}^\delta$ is defined similarly to $A_{cs}$ but with $A+\delta A$ in place of $A$. Then, for all $\delta\in\rline$ sufficiently small, $\Sigma_c$ is a stabilizing output feedback controller for $\Sigma_{cs}^\delta$.
\end{theorem} \vspace{-3mm}
\end{framed}
\vspace{-6mm}

\begin{proof}
Let $A_{cs}'=\sbm{E+LG & (F+LH)C_\L \\ 0 & A}$. Since $A_{cs}'$ has the same triangular structure as $A_{cs}$ with matrices $E+LG$ and $F+LH$ in place of $E$ and $F$, we can conclude using the regularity of $(A,B,C)$ that $A_{cs}'$, like $A_{cs}$, is the generator of a semigroup on $Z_{cs}$ and $B_{cs}$ is an admissible control operator for this semigroup. This and the boundedness of $K_{cs}=[K_1\ \ K_2]$ implies, see discussion below Definition \ref{def:stab}, that $A_c=A_{cs}'+B_{cs}K_{cs}$ is the generator of a semigroup on $Z_{cs}$. Consequently, noting that $B_c$ and $C_c$ are bounded operators, we get that $(A_c,B_c,C_c,D_c)$ are the GOs of a RLS $\Sigma_c$. This RLS can be written as follows: for $t>0$ \vspace{-1mm}
\begin{align}
 \dot{\tilde w}(t) &= (E+LG) \tilde w(t) + (F+LH)C_\L \tilde z(t)-L\tilde u(t), \label{eq:act:obsode}\\[0.5ex]
 \dot{\tilde z}(t) &= (A+BK_2) \tilde z(t) + BK_1\tilde w(t), \label{eq:act:obspde}\\[0.5ex]
 \tilde y(t) &= K_1 \tilde w(t) + K_2 \tilde z(t), \label{eq:act:obsoutput}\\[-4ex] \nonumber
\end{align}
where $[\tilde w(t)\ \ \tilde z(t)]\in Z_{cs}$, $\tilde u(t)\in\rline^p$ and $\tilde y(t)\in\rline^m$ are the state, input and output.

The transfer functions of $\Sigma_{cs}$ and $\Sigma_c$ are $\GGG_{cs}=C_{cs}(sI-A_{cs})^{-1}B_{cs}$ and $\GGG_c=C_c(sI-A_c)^{-1}B_c$. Since $B_c$ and $C_c$ are bounded it follows using \eqref{eq:Best} or \eqref{eq:Cest} that $\lim_{\Re s\to\infty} \|\GGG_c(s)\|_{\Lscr(\rline^p,\rline^m)}=0$. Therefore $\lim_{\Re s\to\infty} \|\GGG_c(s) \GGG_{cs}(s)\|_{\Lscr(\rline^m)}=0$ and so $I$ is an admissible feedback operator for $\GGG_c\GGG_{cs}$. Clearly, $I-D_cD_{cs}$ is invertible. Hence the positive feedback interconnection of $\Sigma_{cs}$ and $\Sigma_c$  (i.e.  $\Sigma_1=\Sigma_{cs}$ and $\Sigma_2=\Sigma_c$ in Figure 1) is a RLS denoted as $\Sigma_{fb}$. Thus for each initial state $[w(0) \ \ z(0)]^\top$ of \eqref{eq:act:ode}-\eqref{eq:act:pde} and $[\tilde w(0)\ \ \tilde z(0)]^\top$ of \eqref{eq:act:obsode}-\eqref{eq:act:obspde}, there exist unique state trajectories $[w \ \ z]^\top$ of \eqref{eq:act:ode}-\eqref{eq:act:pde} and $[\tilde w\ \ \tilde z]^\top$ of \eqref{eq:act:obsode}-\eqref{eq:act:obspde} with $\tilde u=G w+HC_\L z$ and $u=K_1\tilde w+K_2\tilde z$. We will prove the exponential stability of $\Sigma_{fb}$ by showing that \vspace{-1mm}
\begin{equation} \label{eq:act:plantobs}
 \|[w(t)\ \ z(t) \ \ \tilde w(t) \ \ \tilde z(t)]^\top \|_{Z_{cs} \times Z_{cs}} \leq Me^{-\o t} \|[w(0) \ \ z(0) \ \ \tilde w(0) \ \ \tilde z(0)]^\top\|_{Z_{cs}\times Z_{cs}} \vspace{-1mm}
\end{equation}
for some $M,\o>0$ and all $t\geq0$. Define $e_w=\tilde w-w$ and $e_z=\tilde z-z$. Then from \eqref{eq:act:ode}, \eqref{eq:act:pde}, \eqref{eq:act:obsode} and \eqref{eq:act:obspde} we get that for almost all $t\geq0$, \vspace{-2mm}
\begin{equation} \label{eq:act:wzewez}
 \bbm{\dot {\tilde w}(t)\\ \dot {\tilde z}(t) \\ \dot e_w(t)\\ \dot e_z(t)} =  \bbm{E & FC_\L & LG & LHC_\L \\ BK_1 & A+BK_2 & 0 & 0 \\ 0 & 0 & E+LG & (F+LH)C_\L\\ 0 & 0 & 0 & A} \bbm{\tilde w(t) \\ \tilde z(t) \\ e_w(t) \\ e_z(t)}. \vspace{-2mm}
\end{equation}
Observe that $A_{cs}+B_{cs}K_{cs}=\sbm{E & FC_\L \\ BK_1 & A+BK_2}$ is exponentially stable, see discussion below \eqref{eq:act:wzdec}, and the exponential stability of $E+LG$ and $A$ imply that $A_{cs}=\sbm{E+LG & (F+LH)C_\L \\ 0 & A}$ is also exponentially stable. It now follows, using \cite[Lemma 5.1]{WeCu:97}, that the semigroup generated by the state operator in \eqref{eq:act:wzewez} is exponentially stable and there exist $M_0,\o>0$ such that for all $t\geq0$, \vspace{-1mm}
$$ \|[w(t)\ \ z(t) \ \ e_w(t) \ \ e_z(t)]^\top \|_{Z_{cs} \times Z_{cs}} \leq M_0e^{-\o t} \|[w(0) \ \ z(0) \ \ e_w(0) \ \ e_z(0)]^\top\|_{Z_{cs}\times Z_{cs}}. \vspace{-1mm} $$
The estimate in \eqref{eq:act:plantobs} follows and therefore $\Sigma_c$ is a stabilizing output feedback controller for $\Sigma_{cs}$.

Next we will establish the robustness claim in the theorem. For each $\delta\in(-1,\infty)$, using the arguments presented above \eqref{eq:act:plantobs}, we get that the feedback interconnection of $\Sigma_{cs}^\delta$ and $\Sigma_c$ is a RLS, denoted as $\Sigma_{fb}^\delta$. So for each initial state $[w(0) \ \ z(0)]^\top$ of \eqref{eq:act:Rode}-\eqref{eq:act:Rpde} and $[\tilde w(0)\ \ \tilde z(0)]^\top$ of \eqref{eq:act:obsode}-\eqref{eq:act:obspde}, there exist unique state trajectories $[w \ \ z]^\top$ of \eqref{eq:act:Rode}-\eqref{eq:act:Rpde} and $[\tilde w\ \ \tilde z]^\top$ of \eqref{eq:act:obsode}-\eqref{eq:act:obspde} with $\tilde u=G w+HC_\L z$ and $u=K_1\tilde w+K_2\tilde z$. We will prove the exponential stability of $\Sigma_{fb}^\delta$ for small $\delta$ by proving that these state trajectories satisfy \eqref{eq:act:plantobs} for some $M,\o>0$. Let $z_a$ be the state trajectory of \vspace{-1mm}
\begin{equation} \label{eq:act:auxstate}
 \dot z_a(t) = A z_a(t) + BK_1 \tilde w(t) + BK_2 \tilde z(t), \qquad z_a(0)=z(0).
\end{equation}
Write $\tilde w$ as $[\tilde w_1\ \ \tilde w_2]^\top$, where $\tilde w_1\in \rline^{n_1}$ and $\tilde w_2\in\rline^{n_2}$. Recall $K$, $\Pi$, $X$, $A_1$ and $B_1$ from Theorem \ref{th:act:stab}. Define $\tilde p_1 = \tilde w_1 + \Pi \tilde z_1$, $\tilde z_1 = [\tilde w_2\ \ \tilde z]^{\top}$,$e_w=\tilde w-w$, $e_z=\tilde z-z_a$, $q_1 = [\tilde p_1 \ \ \tilde z_1 \ \ e_w\ \ e_z]^\top$ and $q_2 = [z\ \ z_a]$. Define
$$ \Ascr_1=\bbm{E_1+\Pi B_1 K & 0 & L_1 G & L_1HC_\L \\ B_1 K & A_1 & 0 & 0 \\ 0 & 0 & E+LG & (F+LH)C_\L \\ 0 & 0 & 0 & A}, \qquad \Bscr_1= \bbm{L_1H\\0\\ F+LH\\0},$$
$$ \Cscr_1 = \bbm{K & 0 & 0 & 0}, \quad \Ascr_2^\delta=\bbm{A+\delta A & 0\\ 0 & A}, \quad \Bscr_2= \bbm{B\\B},\quad \Cscr_2 = \bbm{-C_\L & C_\L}.$$
Then from \eqref{eq:act:Rode}, \eqref{eq:act:Rpde}, \eqref{eq:act:obsode}, \eqref{eq:act:obspde} and \eqref{eq:act:auxstate} it follows that for almost all $t\geq0$
\begin{equation} \label{eq:act:extended}
 \bbm{\dot q_1(t) \\ \dot q_2(t)} = \bbm{\Ascr_1 & \Bscr_1\Cscr_2 \\ \Bscr_2\Cscr_1 & \Ascr_2^\delta} \bbm{q_1(t) \\ q_2(t)}.
\end{equation}
Since $\sbm{E_1+\Pi B_1 K & 0 \\ B_1 K & A_1 }$ and $\sbm{E+LG & (F+LH)C_\L \\ 0 & A}$ are exponentially stable, $\Ascr_1$ is the generator of an exponentially stable semigroup on $V=\rline^{n_1}\times X \times \rline^n\times Z$. (In fact, $\Ascr_1$ and the state operator in \eqref{eq:act:wzewez} are similar via a bounded transformation.) Clearly $\Bscr_1\in\Lscr(\rline^q,V)$ and $\Cscr_1\in\Lscr(V,\rline^m)$. From these it follows that $(\Ascr_1,\Bscr_1,\Cscr_1,0)$ are the GOs of an exponentially stable RLS $\Sigma_1$.
From the regularity of $(A,B,C)$ and Assumption \ref{as:act:exp}, it follows that $(\Ascr_2^\delta,\Bscr_2,\Cscr_2,0)$ are the GOs of an exponentially stable RLS $\Sigma_2^\delta$. The transfer function $\GGG_1$ of $\Sigma_1$ is in $H^\infty(\Lscr( \rline^q, \rline^m))$ and for all $s\in\overline{\cline^+}$,
$$ \GGG_1(s)=K(sI-E_1-\Pi B_1K)^{-1}L_1[G(sI-E-LG)^{-1}(F+LH)+H]. $$
The transfer function $\GGG_2^\delta$ of $\Sigma_2^\delta$ is in $H^\infty(\Lscr(\rline^m,\rline^q))$ and for all $s\in\overline{\cline^+}$,
\begin{align*}
 \GGG_2^\delta(s) &= C_\L(sI-A)^{-1}B - C_\L(sI-A-\delta A)^{-1}B \\
 &= \delta C_\L(sI-A-\delta A)^{-1}B - \delta s C_\L(sI-A)^{-1}(sI-A-\delta A)^{-1}B.
\end{align*}
Since all the operators (matrices) in the expression for $\GGG_1$ are bounded, it follows that $\lim_{|s|\to\infty,\, s\in \overline{\cline^+}} \|\GGG_1(s)\|=0$. From the expression for $\GGG_2^\delta$, using \eqref{eq:Best} and \eqref{eq:Cest}, we have $\lim_{\delta\to 0}\sup_{s\in S} \|\GGG_2^\delta(s)\|=0$ for any compact subset $S$ of $\overline{\cline^+}$ and, furthermore, $\sup_{\delta\in \Delta} \|\GGG_2^\delta\|_{H^\infty}<\infty$ for any compact subset $\Delta$ of $(-1,\infty)$. Consequently, for all $\delta$ sufficiently small, $\|\GGG_1\GGG_2^\delta\|_{H^\infty}<1$ and so $(I-\GGG_1 \GGG_2^\delta)^{-1} \in H^\infty(\Lscr(\rline^m))$. Thus the positive feedback interconnection of $\Sigma_1$ and $\Sigma_2^\delta$ is an exponentially stable RLS \cite[Proposition 4.6]{WeCu:97} and its state operator is the state operator in \eqref{eq:act:extended}. So the state trajectory $[q_1\ \ q_2]$ of \eqref{eq:act:extended} converges to zero exponentially, implying that the state trajectories $[w \ \ z]^\top$ of \eqref{eq:act:Rode}-\eqref{eq:act:Rpde} and $[\tilde w\ \ \tilde z]^\top$ of \eqref{eq:act:obsode}-\eqref{eq:act:obspde} satisfy \eqref{eq:act:plantobs} for some $M,\o>0$. Hence $\Sigma_{fb}^\delta$ is exponentially stable, i.e. $\Sigma_c$ is a stablizing output feedback controller for $\Sigma_{cs}^\delta$ for small $\delta$.
\end{proof}

The following remark discusses how the controller design techniques proposed in this section can be applied to the PDE-ODE cascade system \eqref{eq:act:ode}-\eqref{eq:act:output} when $B\in\Lscr(\rline^m, Z_{-1})$ is not an admissible control operator for $\tline$.

\begin{remark} \label{rm:act:nonreg}
In the PDE-ODE cascade system \eqref{eq:act:ode}-\eqref{eq:act:output}, suppose that the control operator $B\in\Lscr(\rline^m,Z_{-1})$ is not admissible for $\tline$. However, let $\GGG$ as defined in \eqref{eq:act:RLStf} exist and be bounded on $\cline^+_\o$ for each $\o>\o_\tline$. Let Assumptions \ref{as:act:exp}, \ref{as:act:stab} and \ref{as:act:det} hold. To apply Theorems \ref{th:act:stab} and \ref{th:act:det} to \eqref{eq:act:ode}-\eqref{eq:act:output}, introduce a stable first-order filter in cascade with the PDE system \eqref{eq:act:pde}, i.e. $u$ in \eqref{eq:act:pde} is obtained as follows: \vspace{-1mm}
\begin{equation} \label{eq:act:Bfilt}
 \dot x_u(t) = -x_u(t) + v(t), \qquad u(t)=x_u(t), \vspace{-1mm}
\end{equation}
where $x_u(t), v(t)\in\rline^m$. Via integration by parts
we get \vspace{-1mm}
\begin{equation*} \label{eq:act:intp}
 \int_0^t \tline_{t-\tau}B x_u(\tau)\dd\tau = \tline_t A^{-1}B x_u(0) - A^{-1}Bx_u(t) - \int_0^t \tline_{t-\tau} A^{-1} B (x_u(\tau)-v(\tau)) \dd\tau. \vspace{-1mm}
\end{equation*}
Consider the operators $\Ascr=\sbm{A & B\\ 0 & -I}$, $\Bscr=\sbm{0\\I}$ and $\Cscr=\bbm{C_\L & 0}$. Using the above integral expression it follows that  $\Ascr$ is the generator of a strongly continuous semigroup $\sline$ on $Z\times\rline^m$ defined as $\sline_t=\sbm{\tline_t &\ \ \int_0^t \tline_{t-\tau}B e^{-\tau} \dd\tau \\ 0 & e^{-t} I}$ for $t\geq0$. Since $\Bscr$ is bounded, it is an admissible control operator for $\sline$. Since $C$ is admissible for $\tline$, $\Cscr$ is an admissible observation operator for $\sline$.  Furthermore, $\Gscr(s)=\Cscr_\L(sI-\Ascr)^{-1} \Bscr=\GGG(s)/(s+1)$ and so $(\Ascr,\Bscr,\Cscr)$ is a regular triple. Consider the PDE-ODE cascade system \eqref{eq:act:ode}-\eqref{eq:act:output} along with the filter \eqref{eq:act:Bfilt}. This system can be written (with input $v$) as \vspace{-1mm}
\begin{align}
 \dot w(t) &= E w(t) + F\Cscr_\L z_c(t), \label{eq:act:odeB} \\[0.5ex]
 \dot z_c(t) &= \Ascr z_c(t) + \Bscr v(t), \label{eq:act:pdeB} \\[0.5ex]
 y(t) &= G w(t) + H \Cscr_\L z_c(t), \label{eq:act:outputB}
\end{align}
where $z_c(t)=[z(t)\ \ x_u(t)]^\top$. The PDE-ODE cascade system \eqref{eq:act:odeB}-\eqref{eq:act:outputB} satisfies all the hypothesis stated in the beginning of this section. Assumptions \ref{as:act:exp}, \ref{as:act:stab} and \ref{as:act:det} also hold for it (this follows from the fact that they hold for \eqref{eq:act:ode}-\eqref{eq:act:output}). Applying Theorems \ref{th:act:stab} and \ref{th:act:det} we obtain state feedback and output feedback controllers which stabilize \eqref{eq:act:odeB}-\eqref{eq:act:outputB}. Clearly, the cascade interconnection of any of these controllers with the filter \eqref{eq:act:Bfilt} is a stabilizing controller for \eqref{eq:act:ode}-\eqref{eq:act:output} (here stabilizing means that the state trajectories of \eqref{eq:act:ode}-\eqref{eq:act:output} in $Z_{cs}$ and the state trajectories of the controller converge to zero exponentially for any initial state). It is also a stabilizing controller for the perturbed system \eqref{eq:act:Rode}-\eqref{eq:act:Rpde}, \eqref{eq:act:output} (this follows via small changes to the robustness arguments in the proof of Theorems \ref{th:act:stab}, \ref{th:act:det}). \hfill$\square$
\end{remark}

In \cite{Kri:2009} and \cite{SaGaKr:2018}, the actuator is modeled as a 1D diffusion equation with Dirichlet boundary control. This model can be written as an abstract linear system with state space $L^2(0,1)$, input space $\rline$ and output space $\rline$. Its state, control, observation and feedthrough operators are defined as follows: $A=\frac{\partial^2}{\partial x^2}$ with $D(A)=\{f\in H^2(0,1) \big| f'(0)=0, f(1)=0\}$, $B=\delta'(1)$ (derivative of Dirac pulse at $x=1$), $Cz=z(0)$ for all $z\in D(A)$ and $D=0$. Its transfer function is $\GGG(s)=1/\cosh(\sqrt s)$. These operators satisfy the hypothesis in Remark \ref{rm:act:nonreg}. Hence for the PDE-ODE cascade systems in \cite{Kri:2009} and \cite{SaGaKr:2018}, stabilizing controllers can be designed using the approach described in the remark.

Suppose \eqref{eq:act:ode} has an additional term $Ju$, i.e. the plant dynamics is governed by 
\begin{equation} \label{eq:act:Jode}
 \dot w(t) = E w(t) + FC_\L z(t) + Ju(t), \vspace{-1mm}
\end{equation}
where $J\in\rline^{n\times m}$. Let $[J_1 \ \ J_2]^\top$ be the partitioning of $J$ corresponding to \eqref{eq:act:Edecom}.
\begin{assumption} \label{as:act:stabnotcas}
$v^\top F_1\GGG(\l)+v^\top J_1\neq0$ for each eigenvalue $\l\in\sigma(E_1)$ and nonzero vector $v\in\rline^{n_1}$ satisfying $v^\top E_1=\l v^\top$.
\end{assumption}

\begin{remark} \label{rm:act:notcas}
Theorem \ref{th:act:stab}, Proposition \ref{pr:act:stabcon} and Theorem \ref{th:act:det} continue to hold if we replace \eqref{eq:act:ode} and \eqref{eq:act:Rode} with \eqref{eq:act:Jode}, Assumption \ref{as:act:stab} with Assumption \ref{as:act:stabnotcas}, $\Pi B_1$ with $\Pi B_1+J_1$ and let $B_{cs} = \sbm{J\\ B}$, $B_1 = \sbm{J_2\\ B}$, $A_c = \sbm{E+LG+JK_1 & (F+LH)C_\L+JK_2 \\ BK_1 & A+B K_2}$. This claim can be proved easily by mimicking the proofs in this section. Hence the results in this section can be used to construct stabilizing controllers for the RLS described by \eqref{eq:act:Jode}, \eqref{eq:act:pde} and \eqref{eq:act:output}. This  remark is useful when Assumption \ref{as:act:exp} does not hold, but the unstable subspace of $A$ is finite-dimensional, see Example 5.1. \hfill $\square$
\end{remark}

\section{ODE plant with PDE sensor}
\label{sec4} \setcounter{equation}{0} 

\ \ \ Consider an ODE-PDE cascade system in which the output of the ODE system drives the PDE system. The ODE models the plant dynamics, while the PDE models the sensor dynamics. The state dynamics of the cascade system is described by the following differential equations: for $t>0$
\vspace{-2mm}
\begin{align}
 \dot w(t) &= E w(t) + F u(t), \label{eq:sen:ode}\\[0.5ex]
 \dot z(t) &= A z(t) + B(G w(t)+H u(t)), \label{eq:sen:pde}\\[-4.5ex] \nonumber
\end{align}
where $w(t)\in\rline^n$ is the plant state, $z(t)\in Z$ is the sensor state, $Z$ is a Hilbert space, $u(t)\in\rline^m$ is the input, $E\in \rline^{n\times n}$ is as in \eqref{eq:act:Edecom}, $F\in \rline^{n\times m}$, $G\in\rline^{q\times n}$, $H\in\rline^{q\times m}$, $A$ is the generator of a strongly continuous semigroup $\tline$ on $Z$ and $B\in \Lscr(\rline^q,Z_{-1})$ is an admissible control operator for $\tline$. The admissibility assumption can be relaxed, see Remark \ref{rm:sen:nonreg}.  The output $y$ of the sensor takes values in $\rline^p$ and is given by \vspace{-2mm}
\begin{equation} \label{eq:sen:output}
 y(t) = C_\L z(t),\qquad t\geq0, \vspace{-2mm}
\end{equation}
where $C\in\Lscr(Z_1,\rline^p)$ is an admissible observation operator for $\tline$. We suppose that the triple $(A,B,C)$ is regular. For the PDE system (sensor), the transfer function $\GGG$ is given in \eqref{eq:act:RLStf}. The combined state space for the plant and sensor is $Z_{cs}=\rline^n\times Z$ and the state, control and observation operators for the combined dynamics (with input $u$, state $[w\ \ z]^\top$ and output $y$) are \vspace{-2.5mm}
\begin{equation*} \label{eq:sen:cas}
 A_{cs} = \bbm{E & 0 \\ BG & A}, \qquad B_{cs} = \bbm{F\\ BH}, \qquad C_{cs} = \bbm{0 & C_\L}. \vspace{-2.5mm}
\end{equation*}
From the feedback theory for RLSs it follows that the cascade system \eqref{eq:sen:ode}-\eqref{eq:sen:output} is a RLS, denoted as $\Sigma_{cs}$, with GOs $(A_{cs}, B_{cs}, C_{cs},0)$, input space $\rline^m$, state space $Z_{cs}$ and output space $\rline^p$.  In Theorem \ref{th:sen:det}, we present an observer for \eqref{eq:sen:ode}-\eqref{eq:sen:output}. This result can be extended easily to a setting in which the output \eqref{eq:sen:output} also contains a term $J w$, see Remark \ref{rm:sen:notcas}. Since the assumptions and results in this section are dual to those in Section \ref{sec3}, we will keep our discussions about them brief.

\begin{assumption} \label{as:sen:exp}
The semigroup $\tline$ (or equivalently $A$) is exponentially stable. \vspace{-1mm}
\end{assumption}

In the context of observer design for \eqref{eq:sen:ode}-\eqref{eq:sen:output}, Assumption \ref{as:sen:exp} is no more restrictive than requiring the pair $(C,A)$ to be detectable. In case this assumption does not hold and the unstable subspace of $A$ is finite-dimensional, we can combine it with the unstable subspace of $E$, redefine $A$, $B$, $C$, $E$, $F$ and $G$ suitably and work with \eqref{eq:sen:ode}-\eqref{eq:sen:output} (with redefined operators) for which Assumption \ref{as:sen:exp} holds, also see Remark \ref{rm:sen:notcas}. Recall the partitioning of $G$ in \eqref{eq:act:part}.

\begin{assumption} \label{as:sen:det}
$\GGG(\l)G_1 v\neq0$ for each eigenvalue $\l\in\sigma(E_1)$ and nonzero vector $v\in\rline^{n_1}$ satisfying $E_1 v=\l v$.
\end{assumption}

When $A$ is exponentially stable, the pair $(C_{cs},A_{cs})$ is detectable if and only if Assumption \ref{as:sen:det} holds, see Proposition \ref{pr:sen:detcon}. When $A$ is not exponentially stable, but $\GGG(\l)$ exists for each $\l\in\sigma(E_1)$, the pair $(C_{cs},A_{cs})$ is detectable if $(C,A)$ is detectable and Assumption \ref{as:sen:det} holds. The next result follows from \cite[Lemma III.4]{NaGiWe:14}. \vspace{-1mm}

\begin{framed} \vspace{-3mm}
\begin{lemma} \label{pr:sen:syl}
Let $\Ascr$ be the generator of an exponentially stable strongly continuous semigroup $\sline$ on a Hilbert space $X$. Let $\Escr\in\rline^{n\times n}$ be such that $\sigma(\Escr)\subset \overline{\cline^+}$. Recall the expression for $e^{-\Escr t}$ from \eqref{eq:act:matexp}. Let $\Bscr\in \Lscr(\rline^n,X_{-1})$. Then  $\Pi\in\Lscr(\rline^n, X)$ defined as \vspace{-2.5mm}
\begin{equation} \label{eq:sen:Pisum}
 \Pi = \sum_{k=1}^v\sum_{j=0}^r (\l_k-\Ascr)^{-1-j} \Bscr E_{kj} \vspace{-2.5mm}
\end{equation}
solves the Sylvester equation \vspace{-3mm}
\begin{equation} \label{eq:sen:syl}
 \Pi \Escr = \Ascr \Pi + \Bscr. \vspace{-2mm}
\end{equation}
\end{lemma} \vspace{-4mm}
\end{framed}
\vspace{-5mm}

\begin{proof}
From the proof of Lemma III.4 in \cite{NaGiWe:14} we get that $\Pi\in\Lscr(\rline^n,X)$ defined as \vspace{-2.5mm}
\begin{equation} \label{eq:sen:NGW}
 \Pi w = \int_0^\infty \sline_t \Bscr e^{-\Escr t}w \dd t \FORALL w\in\rline^n \vspace{-2.5mm}
\end{equation}
solves \eqref{eq:sen:syl}. Substituting for  $e^{-\Escr t}$ from \eqref{eq:act:matexp} into \eqref{eq:sen:NGW} and then using the integral expression for the powers of the resolvent operator, it is easy to verify that $\Pi$ in \eqref{eq:sen:NGW} can equivalently be expressed via the formula in \eqref{eq:sen:Pisum}. \vspace{-1mm}
\end{proof}

We now present an observer for the ODE-PDE cascade system \eqref{eq:sen:ode}-\eqref{eq:sen:output}. Recall the notation $E_1$, $E_2$, $F_1$, $F_2$, $G_1$, $G_2$, $w_1$ and $w_2$ from \eqref{eq:act:Edecom}, \eqref{eq:act:part}. Define \vspace{-1mm} $z_1 = [w_1 \ \ z]^\top$.

\begin{framed} \vspace{-3mm}
\begin{theorem} \label{th:sen:det}
Consider the cascade system \eqref{eq:sen:ode}-\eqref{eq:sen:output}. Suppose that Assumption \ref{as:sen:exp} holds. Define \vspace{-1mm}
$$ A_1 = \bbm{E_2 & 0\\ BG_2 & A}, \qquad B_1 = \bbm{0_{n_2\times q}\\ B}, \qquad C_1 = \bbm{0_{p\times n_2} & C}. \vspace{-1mm}$$
Then $A_1$ is the generator of an exponentially stable strongly continuous semigroup $\sline$ on $X=\rline^{n_2}\times Z$, the control operator $B_1\in\Lscr(\rline^q, X_{-1})$ and the observation operator $C_1\in\Lscr(X_1,\rline^p)$ are admissible for $\sline$ and the triple $(A_1,B_1,C_1)$ is regular. There exists $\Pi\in\Lscr(\rline^{n_1},X)$ such that \vspace{-1mm}
\begin{equation} \label{eq:sen:sylth}
 \Pi E_1 w_1  = A_1\Pi w_1 + B_1G_1 w_1 \FORALL w_1\in\rline^{n_1}
\end{equation}
and $C_{1\L}\Pi\in\Lscr(\rline^{n_1},\rline^p)$.

Suppose that Assumption \ref{as:sen:det} also holds. Then the pair $(C_{1\L}\Pi,E_1)$ is detectable. Fix $L\in \rline^{n_1\times p}$ such that $E_1+LC_{1\L}\Pi$ is Hurwitz. Let $\Pi=[\Pi_1 \ \ \Pi_2]^\top$, where  $\Pi_1\in\Lscr(\rline^{n_1},\rline^{n_2})$ and $\Pi_2\in\Lscr( \rline^{n_1},Z)$. Define $\tilde L = [L \ \ \Pi_1 L]^\top$. Then \vspace{-1mm}
\begin{equation} \label{eq:sen:obs}
 \bbm{\dot{\tilde w} \\ \dot{\tilde z}} = \bbm{E & \tilde L C_\L \\ B G & A+\Pi_2 L C_\L}\bbm{\tilde w \\ \tilde z} - \bbm{\tilde L \\ \Pi_2 L} y + \bbm{F\\B H}u. \vspace{-1mm}
\end{equation}
is an observer for $\Sigma_{cs}$.
\end{theorem} \vspace{-3mm}
\end{framed}

\vspace{-4mm}
\begin{proof}
The exponential stability of the semigroup $\sline$ generated by $A_1$ and the regularity of the triple $(A_1,B_1,C_1)$ can be established like in the proof of Theorem \ref{th:act:stab}. Since $B_1$ is admissible for $\sline$, so is $B_1G_1$. Applying Lemma \ref{pr:sen:syl} with $\Escr=E_1$, $\Ascr=A_1$ and $\Bscr=B_1G_1$, we get that there exists a $\Pi\in\Lscr(\rline^{n_1},X)$ which solves \eqref{eq:sen:sylth}. It follows from  the regularity of the triple $(A_1,B_1,C_1)$ and the expression for $\Pi$ in \eqref{eq:sen:Pisum} that $C_{1\L}\Pi\in\Lscr(\rline^{n_1},\rline^p)$.

Suppose that Assumption \ref{as:sen:det} holds. Then the pair $(C_{1\L}\Pi,E_1)$ is detectable. Indeed, if not, then via the Hautus test there exists a $\l\in\sigma(E_1)$ and a non-zero $v\in\rline^{n_1}$ such that \vspace{-1mm}
\begin{equation} \label{eq:sen:hau}
  E_1 v= \l v, \qquad C_{1\L}\Pi v=0. \vspace{-1mm}
\end{equation}
Choosing $w_1=v$ in \eqref{eq:sen:sylth} and then applying $C_{1\L}(\l I-A_1)^{-1}$ from the left to both sides of the resulting expression,  we get using the first expression in \eqref{eq:sen:hau} and $C_{1\L}(\l I-A_1)^{-1}B_1=\GGG(\l)$ that \vspace{-1mm}
\begin{equation} \label{eq:sen:hausyl}
  C_{1\L}\Pi v = \GGG(\l)G_1 v . \vspace{-1mm}
\end{equation}
Using the second expression in \eqref{eq:sen:hau} it follows from \eqref{eq:sen:hausyl} that $\GGG(\l)G_1 v =0$, which contradicts Assumption \ref{as:sen:det}. Hence the pair $(C_{1\L}\Pi,E_1)$ is detectable.

Fix $L\in \rline^{n_1\times p}$ such that $E_1+LC_{1\L}\Pi$ is Hurwitz. As in the statement of the theorem, let $\Pi=[\Pi_1 \ \ \Pi_2]^\top$ and $\tilde L = [L \ \ \Pi_1 L]^\top$. Define $L_{cs}=[\tilde L\ \ \Pi_2 L]^\top\in\Lscr(\rline^p,Z_{cs})$. Since $L_{cs}$ is bounded, it is an admissible control operator for the semigroup generated by $A_{cs}$ and $\GGG_{L}(s)=C_{cs,\L} (sI-A_{cs})^{-1} L_{cs}$ exists for all $s\in\rho(A_{cs})$. From \eqref{eq:Cest}, $\lim_{\Re s\to\infty} \|\GGG_{L}(s)\|_{\Lscr(\rline^p)}=0$, which implies that $(A_{cs}, L_{cs}, C_{cs})$ is a regular triple and $I$ is an admissible feedback operator for $\GGG_{L}$. To establish that \eqref{eq:sen:obs} is an observer for $\Sigma_{cs}$, according to Definition \ref{def:det} and the discussion below it, we only need to show that $A_{cs}+L_{cs}C_{cs}$ is exponentially stable, i.e. for each $[e_w(0)\ \ e_z(0)]^\top\in Z_{cs}$ the state trajectory of \vspace{-2mm}
\begin{equation} \label{eq:sen:ewez}
 \bbm{\dot e_w(t)\\ \dot e_z(t)} = \bbm{E  & \tilde L C_\L \\  B G & A+\Pi_2L C_\L}\bbm{e_w(t)\\e_z(t)} \vspace{-1mm}
\end{equation}
satisfies the following estimate for some $M,\o>0$: \vspace{-1mm}
\begin{equation} \label{eq:sen:ewezdec}
  \|e_w(t)\|+\|e_z(t)\| \leq M e^{-\o t}(\|e_w(0)\|+\|e_z(0)\|) \FORALL t\geq0.
\end{equation}
Let $e_w=[e_{w1}\ \ e_{w2}]^\top$ with $e_{w1}\in\rline^{n_1}$ and $e_{w2}\in\rline^{n_2}$. Define $e_{z1}=[e_{w2}\ \ e_z]^\top-\Pi e_{w1}$. Then along the trajectory of \eqref{eq:sen:ewez} we get that for almost all $t\geq0$
$$  \bbm{\dot e_{w1}(t)\\ \dot e_{z1}(t)} = \bbm{E_1+LC_{1\L}\Pi  & L C_{1\L} \\ 0 & A_1}\bbm{e_{w1}(t) \\ e_{z1}(t)}. $$
From the exponential stability of $E_1+LC_{1\L}\Pi$ and $A_1$ and the upper triangular form of the state operator, we get that $ \|e_{w1}(t)\|+\|e_{z1}(t)\| \leq M_1 e^{-\o t}(\|e_{w1}(0)\|+\|e_{z1}(0)\|)$ for some $M_1,\o>0$ and all $t\geq0$, from which \eqref{eq:sen:ewezdec} follows. \vspace{-2mm}
\end{proof}

Theorem \ref{th:sen:det} shows that Assumption \ref{as:act:stab} is sufficient for the existence of an observer for the ODE-PDE system \eqref{eq:sen:ode}-\eqref{eq:sen:output}. The next proposition establishes that this assumption is also necessary. \vspace{-1mm}

\begin{framed} \vspace{-3mm}
\begin{proposition} \label{pr:sen:detcon}
Consider the cascade system \eqref{eq:sen:ode}-\eqref{eq:sen:output}. Let Assumption \ref{as:sen:exp} hold. Then the pair $(C_{cs}, A_{cs})$ is detectable if and only if Assumption \ref{as:sen:det} holds.
\end{proposition} \vspace{-3mm}
\end{framed}

\vspace{-5mm}
\begin{proof}
Suppose that Assumption \ref{as:sen:det} holds. We have shown in Theorem \ref{th:sen:det} that $(C_{cs},A_{cs})$ is detectable and found $L_{cs}$ such that $A_{cs}+L_{cs}C_{cs}$ is exponentially stable.

Conversely, suppose that the pair $(C_{cs},A_{cs})$ is detectable. If Assumption \ref{as:sen:det} does not hold, then there exists a non-zero $v\in\rline^{n_1}$ such that $E_1 v=\l v$ for some $\l\in\sigma(E_1)$ and $\GGG(\l)G_1 v=0$. It now follows from \eqref{eq:sen:hausyl} that $C_{1\L} \Pi v=0$. Define $V=[v\ \ \Pi v]^\top$. Noting that $A_{cs} = \sbm{E_1 & 0 \\ B_1G_1 & A_1}$ and $C_{cs} = \bbm{0 & C_{1\L}}$, it is easy to verify using \eqref{eq:sen:sylth} that $V\in \Dscr(A_{cs})$, $A_{cs}V=\l V$ and $C_{cs}V=0$. Hence for any $L_{cs}\in \Lscr(\rline^p,Z_{cs,-1})$ we have $(A_{cs}+L_{cs}C_{cs}) V=\l V$ which, along with $\Re\lambda\geq0$, implies that $A_{cs}+L_{cs}C_{cs}$ is not exponentially stable, which in turn contradicts the detectability of the pair $(C_{cs},A_{cs})$. Hence Assumption \ref{as:sen:det} must hold. \vspace{-1mm}
\end{proof}

The next remark discusses the construction of an observer for \eqref{eq:sen:ode}-\eqref{eq:sen:output} when the control operator $B\in\Lscr(\rline^q,Z_{-1})$ is not admissible for $\tline$.

\begin{remark} \label{rm:sen:nonreg}
In the cascade system \eqref{eq:sen:ode}-\eqref{eq:sen:output}, suppose that  $B\in\Lscr(\rline^q,Z_{-1})$ is not admissible for $\tline$. However, let $\GGG$ in \eqref{eq:act:RLStf} exist and be bounded on $\cline^+_\o$ for each $\o>\o_\tline$ and let Assumptions \ref{as:sen:exp} and \ref{as:sen:det} hold. Then, via arguments similar to those used in Remark \ref{rm:act:nonreg} to show that $(\Ascr,\Bscr,\Cscr)$ is a regular triple, we can establish that $A_1$ is the generator of an exponentially stable semigroup and $(A_{cs},L_o,C_{cs})$ is a regular triple for any $L_o\in\Lscr(\rline^p,Z_{cs})$ (the role of the first-order filter in the arguments in Remark \ref{rm:act:nonreg} will be played by the ODE system in the arguments here). Clearly $B_1\in\Lscr(\rline^q, \rline^{n_2}\times Z_{-1})$ and $C_{1\L}(sI-A_1)^{-1}B_1$ (being equal to $\GGG(s)$) exists if $\Re s>\o_\tline$. Let $\Pi$ solve \eqref{eq:sen:sylth} and define $L_{cs}$ as in the proof of Theorem \ref{th:sen:det}. Then like in that proof we can show that $I$ is an admissible feedback operator for $C_{cs,\L} (sI-A_{cs})^{-1} L_{cs}$ and $A_{cs}+L_{cs}C_{cs}$ is exponentially stable, i.e. the pair $(C_{cs},A_{cs})$  is detectable. In addition, if $H=0$, then
\eqref{eq:sen:obs} is an observer for \eqref{eq:sen:ode}-\eqref{eq:sen:output}. \hfill$\square$
\end{remark}

The sensor model in \cite{Kri:2009} is the 1D diffusion equation described below Remark \ref{rm:act:nonreg}, and for it all the hypothesis in the above remark (including $H=0$) hold.

Suppose that we modify \eqref{eq:sen:output} to include an additional term $Jw$, i.e. \vspace{-1.5mm}
\begin{equation} \label{eq:sen:Joutput}
 y(t) = C_\L z(t) + Jw(t), \vspace{-1.5mm}
\end{equation}
where $J\in\rline^{p\times n}$. Let $[J_1 \ \ J_2]$ be the partitioning of $J$ corresponding to \eqref{eq:act:Edecom}.
\begin{assumption} \label{as:sen:detnotcas}
$C_\L(\l I -A)^{-1} B G_1 v+ J_1 v\neq0$ for each eigenvalue $\l\in\sigma(E_1)$ and nonzero vector $v\in\rline^{n_1}$ satisfying $E_1 v=\l v$.
\end{assumption}

\begin{remark} \label{rm:sen:notcas}
Theorem \ref{th:sen:det} and Proposition \ref{pr:sen:detcon} continue to hold if we replace \eqref{eq:sen:output} with \eqref{eq:sen:Joutput} provided we replace Assumption \ref{as:sen:det} with Assumption \ref{as:sen:detnotcas}, $C_{1\L}\Pi$ with $C_{1\L}\Pi+J_1$ and let
$C_{cs}=[J\ \ C]$ and $C_1=[J_2\ \ C]$ and change $\sbm{E & \tilde L C_\L \\ B G & A+\Pi_2 L C_\L}$ to $\sbm{E+\tilde L J & \tilde L C_\L \\ B G+\Pi_2 L J & A+\Pi_2 L C_\L}$. This claim can be established easily by mimicking the proofs in this section. This remark, like Remark \ref{rm:act:notcas}, is useful when Assumption \ref{as:sen:exp} does not hold, but the unstable subspace of $A$ is finite-dimensional. In this case, if we adopt the approach of redefining operators discussed below Assumption \ref{as:sen:exp}, a $Jw$ term will typically appear in \eqref{eq:sen:output} after the redefinition. \hfill$\square$ \vspace{-2mm}
\end{remark}

\section{Illustrative examples} \label{sec5} \setcounter{equation}{0} 
\vspace{-2mm}
\ \ \ In Example 5.1, we illustrate the results in Section \ref{sec3} by constructing a robust output feedback controller for stabilizing an unstable plant driven by an unstable actuator modeled as a 1D diffusion equation. In Example 5.2, we illustrate the results in Section \ref{sec4} by constructing an observer for an unstable plant with a stable sensor modeled as a 1D wave equation. 

\begin{example}
Let the plant \eqref{eq:act:ode} and its output \eqref{eq:act:output} be determined by the matrices \vspace{-2mm}
$$ E = \bbm{0 & 1\\ -1 & 0}, \quad F = \bbm{0 \\ 1}, \quad G = \bbm{1 & 0}, \quad H = 0. \vspace{-1mm} $$
Since $E$ has no stable eigenvalues, $E_1=E$ and $F_1=F$.
Let the actuator dynamics be governed by the diffusion PDE \vspace{-1.5mm}
\begin{align}
 z_t(x,t) &=  z_{xx}(x,t) \FORALL x\in(0,1), \FORALL t>0,\nonumber\\
 z_x(0,t) &= 0, \qquad z_x(1,t)=u(t),\label{eq:ex1:actuator} \\[-4ex]\nonumber
\end{align}
where the function $z(\cdot,t)$ is the state and $u(t)\in\rline$ is the input to the actuator. The plant is driven by the actuator output $z(0,t)\in \rline$. The actuator dynamics can be written as an abstract evolution equation of the form \eqref{eq:act:pde} on the state space $Z=L^2(0,1)$ with state operator $A$ defined as $A\phi= \phi_{xx}$ for all $\phi\in D(A)$, where $D(A) = \{\phi\in H^2(0,1) \big| \phi_x(0)= \phi_x(1)=0\},$ and control operator $B=\delta_1$, where $\delta_1$ is the Dirac pulse at $x=1$. The observation operator $C$ for the actuator output is defined as $C\phi=\phi(0)$ for all $\phi\in D(A)$. The operator $A$ has eigenvalues $\l_n=-n^2\pi^2$, $n\geq0$, with corresponding eigenfunctions $\phi_n(x)=\sqrt{2}\cos n\pi x$ for $n\geq1$, $\phi_0=1$, which form an orthonormal basis in $L^2(0,1)$ \cite[Example 2.3.7]{Cur-Zw}. Hence $A$ is a Riesz spectral operator and it generates a semigroup $\tline$ on $Z$. The admissibility of $B\in\Lscr(U,Z_{-1})$ and $C\in\Lscr(Z_1,\rline)$ for $\tline$ and the regularity of the triple $(A,B,C)$ follow from \cite{BGSW}, see also \cite[Example VI.1]{NaGiWe:14}. The actuator transfer function, see \eqref{eq:act:RLStf}, is $\GGG(s)= 1{\big /}(\sqrt{s}\sinh\sqrt{s})$ for $\Re s >0$. 

While $A$ is not stable, the pair $(A,B)$ is stabilizable. Indeed, $A+B K_\L$ is stable for $K$ defined as $K\phi=-\phi(1)$ for all $\phi\in D(A)$ \cite[Example VI.1]{NaGiWe:14}. Furthermore, $\GGG(\l)$ exists for each $\l\in E_1$ and Assumption \ref{as:act:stab} holds. It follows from the discussions below Assumptions \ref{as:act:exp} and \ref{as:act:stab} that the pair $(A_{cs},B_{cs})$ is stabilizable. The unstable subspace $Z_u$ of $A$ is the span of $\phi_0$ and its stable subspace $Z_s$ is the orthogonal complement of $\phi_0$ in $L^2(0,1)$. Since $Z_u$ is finite-dimensional, as suggested below Assumptions \ref{as:act:exp}, we will combine it with the unstable subspace of $E$, redefine the operators suitably so that Assumptions \ref{as:act:exp} and \ref{as:act:stab} hold for the redefined operators and finally design a stabilizing output-feedback controller for the above interconnection using Theorem \ref{th:act:det} and Remark \ref{rm:act:notcas}.

The restriction of the actuator dynamics in \eqref{eq:ex1:actuator} to $Z_u$, obtained by taking the innerproduct of \eqref{eq:act:pde} with $\phi_0$, is $\dot z_u(t) = u(t)$ and its restriction to  $Z_s$ is $\dot z_s(t) = A_s z_s(t) + B_s u(t)$. Here $A_s$ is the restriction of $A$ to $Z_s$ and $B_s=(B-\phi_0)$. Clearly $A_s$ is exponentially stable and the regularity of the triple $(A_s,B_s,C)$ follows from the regularity of $(A,B,C)$. Combining the unstable part of the actuator dynamics with the plant dynamics, the new finite-dimensional dynamics is given by \eqref{eq:act:Jode} and output is given by \eqref{eq:act:output}, where \vspace{-2mm}
$$ E = \bbm{0 & 1 & 0\\ -1 & 0 & 1 \\ 0 & 0 & 0}, \quad F = \bbm{0\\ 1\\ 0}, \quad J =\bbm{0\\ 0\\ 1}, \quad G = \bbm{1 & 0 & 0}, \quad H = 0. \vspace{-2mm} $$
This dynamics is driven by the stable part of the actuator dynamics which, after redefining $A$ and $B$ to be $A_s$ and $B_s$, is given by \eqref{eq:act:pde}. Clearly, for the redefined operators, $E_1=E$, $F_1=F$, $J_1=J$, $G_1=G$, $A_1=A$, $B_1=B$ and $C_1=C$, Assumption \ref{as:act:exp} holds and, since $(A_{cs},B_{cs})$ is stabilizable, Assumption \ref{as:act:stabnotcas} must also hold according to Proposition \ref{pr:act:stabcon} and Remark \ref{rm:act:notcas}. It is easy to verify that Assumption \ref{as:act:det} is satisfied. We will apply Theorem \ref{th:act:det}, taking into account Remark \ref{rm:act:notcas}, to design a robust stabilizing output feedback controller. In what follows, we work with the redefined operators.

Using Lemma \ref{pr:act:syl}, \eqref{eq:act:matexp} and \eqref{eq:act:Pisum}, it follows after a simple calculation that \vspace{-2mm}
\begin{equation} \label{eq:ex1:Pi}
 \Pi= \frac{1}{2}\bbm{i\\ 1 \\ 0 }C_\L (-iI-A)^{-1} + \frac{1}{2}\bbm{-i\\ 1 \\ 0}C_\L (iI-A)^{-1} \vspace{-2mm}
\end{equation}
solves \eqref{eq:act:sylth}. From the Riesz spectral property of $A$ we have $(\l I-A)^{-1}z=\sum_{n=1}^\infty \frac{\langle z, \phi_n \rangle}{\l-\l_n}\phi_n$ for all $z\in Z_s$ and $\l\in \rho(A)$. This series converges in $Z_1$. Hence we can compute $C(\l I-A)^{-1}z$ by applying $C$ to each term of the series. Using this it follows from \eqref{eq:ex1:Pi} that \vspace{-4mm}
\begin{equation} \label{eq:ex1:Piseries}
 \Pi z =  -\sqrt{2} \sum_{n=1}^\infty \frac{\langle z, \phi_n \rangle}{1+\l_n^2}\bbm{1\\ \l_n\\0}. \vspace{-2mm}
\end{equation}
Noting that $C_\L(sI-A)^{-1}B=\GGG(s)-1/s$, we get from \eqref{eq:ex1:Pi} after a simple calculation that $ \Pi B = \bbm{0.019 & -0.165 & 0}^\top.$ Let $K=\bbm{2.522&-1.361&-3.273}$ and $L=\bbm{-3 & -1.75 & -0.75}^\top$ so that $E+(\Pi B+J)K$ and $E+LG$ are Hurwitz. By definition $K_1=K$ and $K_2=K\Pi$. The RLS $\Sigma_c$ with GOs $(A_c,B_c,C_c, D_c)$, where $B_c$, $C_c$ and $D_c$ are as in Theorem \ref{th:act:det} and $A_c$ is as in Remark \ref{rm:act:notcas}, is the required robust stabilizing output feedback controller. We have validated this controller by implementing the closed-loop of the actuator-plant cascade system and the controller numerically. In our simulation, the initial condition for the plant is $[1\ \ 1]^\top$. All the other initial conditions are zero. To implement $K_2$, we approximate $\Pi$ by truncating the series in \eqref{eq:ex1:Piseries} after 10 terms. Figure 2 shows the plant state trajectory.

\m\vspace{-14mm}
$$\includegraphics[scale=0.9]{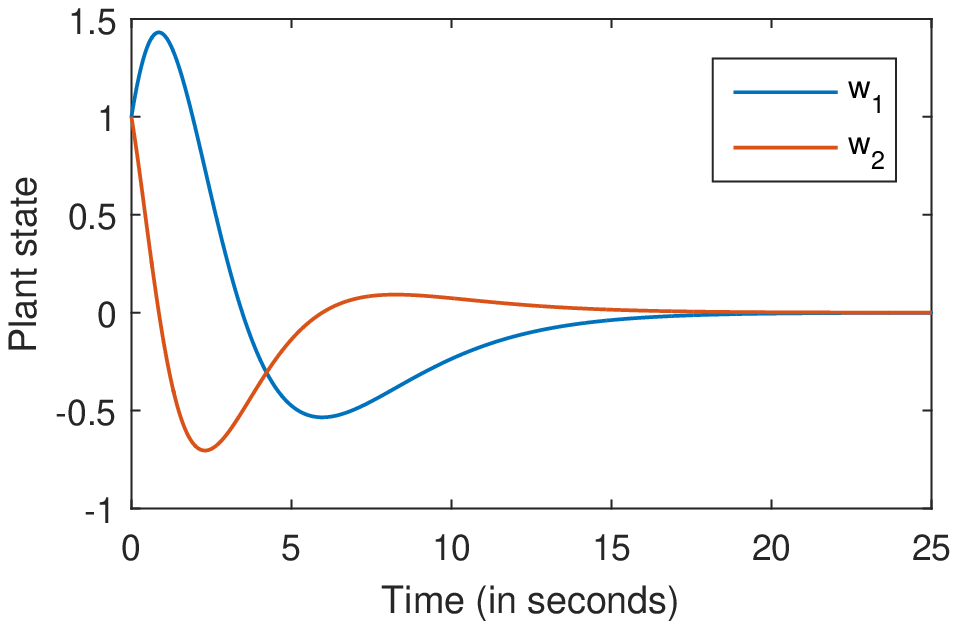}$$
\centerline{ \parbox{5.7in}{\vspace{1mm}
   Figure 2. The controller designed for the actuator-plant cascade system in Example 5.1 ensures that plant state $w=[w_1\ \ w_2]^\top$ converges to zero exponentially. \vspace{-2mm}}}
\end{example}

\begin{example}
Let the plant in \eqref{eq:sen:ode} be determined by the matrices $E$ and $F$ defined in Example 5.1. The plant output which drives the sensor is $\bar Gw$, where $\bar G=[1\ \ 0]$. Let the sensor dynamics be governed by the wave PDE \vspace{-2mm}
\begin{align}
 \bar z_{tt}(x,t) &= \bar z_{xx}(x,t) \FORALL x\in(0,1), \FORALL t>0,\nonumber\\[0.5ex]
 \bar z_x(0,t) &= \bar z_t(0,t), \qquad \bar z(1,t)=\bar G w(t). \label{eq:ex2:sensor} \\[-4.8ex]\nonumber
\end{align}
The sensor output is $\bar z(0,t)$. A similar sensor model is considered in \cite{Kri:2009a}, where the stabilizing term $\bar z_t(0,t)$ is a part of the observer rather than the sensor model. In both cases, the resulting observer error dynamics to be stabilized is the same. It is difficult to formulate the above sensor dynamics directly as an abstract evolution equation. Hence we introduce the transformation $z(x,t)=\bar z(x,t)-x^2 \bar G w(t)$. Then $z$ satisfies the wave PDE \vspace{-2.5mm}
\begin{align}
 z_{tt}(x,t) &= z_{xx}(x,t)+(2\bar G-x^2\bar GE^2)w(t) - x^2\bar GEF u(t) \quad \forall x\in(0,1), \quad \forall\ t>0,\nonumber\\[0.5ex]
 z_x(0,t) &= z_t(0,t), \qquad z(1,t)=0, \label{eq:ex2:sensormod}\\[-4.8ex]\nonumber
\end{align}
which we regard as the sensor dynamics for observer design. The output of this sensor is $z(0,t)$, which is the same as $\bar z(0,t)$. The dynamics in \eqref{eq:ex2:sensor} and \eqref{eq:ex2:sensormod} are equivalent under some regularity assumptions on their solutions; such an assumption is implicit in the observer design in \cite{Kri:2009a}. For instance, for any $C^1$ input $u$, $z$ is a classical solution of \eqref{eq:ex2:sensormod} if and only if $\bar z(x,t)=z(x,t)+x^2\bar G w(t)$ is a classical solution of \eqref{eq:ex2:sensor}. Also, the mild solution of \eqref{eq:ex2:sensormod} in $Z=H^1_0(0,1)\times L^2(0,1)$ can be shown to yield a weak solution of \eqref{eq:ex2:sensor}. An observer built for the plant-sensor system by regarding \eqref{eq:ex2:sensormod} as the sensor dynamics is also an observer for the plant-sensor system in which the sensor dynamics is \eqref{eq:ex2:sensor}. To be precise, it will generate exponentially accurate estimates of $w$ and $\bar z(x,t)-x^2 \bar G w(t)$. We illustrate this below in our simulation. \vspace{-1mm}

Let $G=\sbm{2\bar G\\ -\bar G E^2}$ and $H=\sbm{0\\-\bar G E F}$. Let $Z=H^1_0(0,1)\times L^2(0,1)$, where $H^1_0(0,1) = \{f\in H^1(0,1) \big| f(1)=0\}$. Define $A$ by $A\sbm{f \\ g}=\sbm{g \\ f_{xx}}$ for all $(f,g)\in D(A)$, where
$D(A) = \{(f,g)\in H^2(0,1)\cap H^1_0(0,1)\times H^1_0(0,1) \big| f_x(0)= g(0)\}$. Define $B\in\Lscr(\rline^2,Z)$ by $B\sbm{a\\b}=(0,a+b x^2)\in Z$. Define $C\in\Lscr(Z,\rline)$ by $C\sbm{f \\ g}=f(0)$ for all $(f,g)\in Z$. It is well-known that $A$ generates an exponentially stable semigroup $\tline$ on $Z$. Since $B$ and $C$ are bounded, they are admissible for $\tline$ and the triple $(A,B,C)$ is regular. With these operators $A$, $B$, $C$, $G$ and $H$ the sensor dynamics \eqref{eq:ex2:sensormod} can be formulated as an abstract evolution equation of the form \eqref{eq:sen:pde} on $Z$ with output \eqref{eq:sen:output}. Next we next design an observer for \eqref{eq:sen:ode}-\eqref{eq:sen:output} determined by the above operators.  \vspace{-1mm}

For each $s\in\overline{\cline^+}$ and $\sbm{f \\ g}\in Z$, we  compute $(sI-A)^{-1}\sbm{f \\ g}$ by solving the ODE $(sI-A)\sbm{\phi \\ \psi}=\sbm{f \\ g}$ to get \vspace{-2mm}
\begin{align}
 &(sI-A)^{-1} \bbm{f \\ g}(x)=\bbm{\phi(x) \\ \psi(x)} \nonumber \\
 &= \bbm{p \cosh sx + \frac{q\sinh sx}{s} - \int_0^x \frac{\sinh s(x-y)}{s}[sf(y)+g(y)]\dd y \\ p s\cosh sx + q\sinh sx - \int_0^x \sinh s(x-y)[sf(y)+g(y)]\dd y -f(x)}, \label{eq:ex2:sI-A} \\[-5ex]\nonumber
\end{align}
where $p$ and $q$ are such that $\phi_x(0)=\psi(0)$ and $\psi(1)=0$. From this expression we get that the sensor transfer function, see \eqref{eq:act:RLStf}, is given by \vspace{-2mm}
$$ \GGG(s)= \bbm{\dfrac{\cosh s -1}{s^2(\sinh s+ \cosh s)} & \dfrac{2\cosh s -2-s^2}{s^4(\sinh s+ \cosh s)}} \FORALL s\in \overline{\cline^+}.\vspace{-2mm} $$
We have $E_1=E$ and so $G_1=G$. It is easy to see that Assumptions \ref{as:sen:exp} and \ref{as:sen:det} hold. Using \eqref{eq:sen:Pisum} and \eqref{eq:ex2:sI-A}, it follows after a lengthy calculation that
$ \Pi= \bbm{\cos(x-1)-x^2 & \sin(x-1) \\ -\sin(x-1) & \cos(x-1)-x^2} $
solves \eqref{eq:sen:sylth}. Clearly $C\Pi = \bbm{\cos 1 & -\sin 1}$. Let $L=\bbm{-2.462 & 1.984}^\top$ so that $E+LC\Pi$ is Hurwitz. Then \eqref{eq:sen:obs} (with $\tilde L = L$, $\Pi_2=\Pi$) is an observer for the plant-sensor system with the sensor dynamics in \eqref{eq:ex2:sensormod}. We have validated this observer numerically on the plant-sensor system in which the sensor dynamics is governed by \eqref{eq:ex2:sensor}. In our simulation, $u(t)=\sin 5t$ in \eqref{eq:sen:ode}. The initial condition for the plant is $[-1\ \ 2]^\top$. All other initial conditions are zero. Figure 3 shows the estimation error in the plant state. \vspace{-6mm}

$$\includegraphics[scale=0.9]{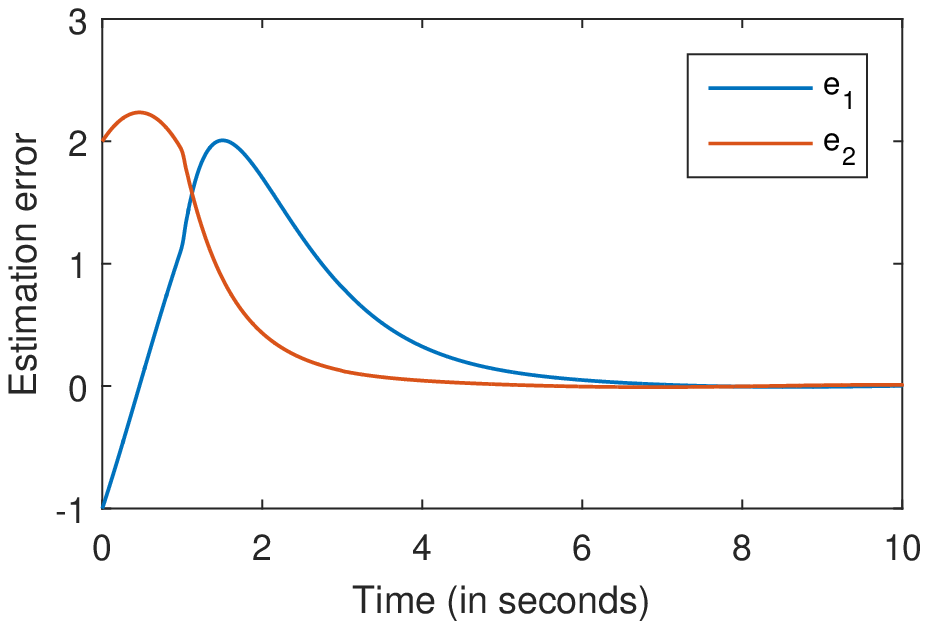}$$
\centerline{ \parbox{5.7in}{\vspace{1mm}
   Figure 3. The error $e_1=w_1-\hat w_1$ and $e_2=w_2-\hat w_2$ between the plant state and its estimate generated by the observer converges to zero exponentially. \vspace{1.5mm}}}
\end{example}

\begin{remark} \label{rm:sylsol}
A key step in the controller/observer design approach presented in this work is solving a Sylvester equation with unbounded operators for $\Pi$ and then computing $\Pi B_1$ (for controller design) or $C_{1\L} \Pi$ (for observer design). These operators can be constructed by first computing the resolvent $(\l I-A_1)^{-1}$ for each $\l\in\sigma(E_1)$ (this follows from the expressions in \eqref{eq:act:Pisum} and \eqref{eq:sen:Pisum}). Also, using the resolvent $(\l I-A)^{-1}$ for each $\l\in\sigma(E_1)$, we can verify the solvability of the stabilization and estimation problems. When the PDE is a 1D system with constant parameters, the resolvent can be computed easily by solving a linear ODE with constant coefficients like in Example 5.2. Developing numerical techniques for computing the resolvent and the operators $\Pi$, $\Pi B_1$ and $C_{1\L} \Pi$ for higher-dimensional PDEs and PDEs with spatially-varying coefficients is a topic for future research. \hfill$\square$ \vspace{-2mm}
\end{remark}

\section{Conclusions and future work}
\label{sec6} \setcounter{equation}{0} 
\vspace{-1mm}

\ \ \ We have presented a Sylvester equation based framework for stabilizing PDE-ODE cascade systems and constructing observers for ODE-PDE cascade systems. Using this framework we can solve the PDE-ODE stabilization and ODE-PDE estimation problems for several PDE models, which have been solved in the literature via the backstepping approach. To be specific, applying Theorem \ref{th:act:stab} we can solve the robust state feedback PDE-ODE stabilization problem considered in \cite{KrSm:2008} for a transport equation, in \cite{Kri:2009} for a diffusion equation (see also Remark \ref{rm:act:nonreg}) and in \cite{Kri:2009a} and \cite{LiKr:2010} for a wave equation. We remark that in the case of the wave equation, we must first stabilize it using the control law in \cite{SmKr:2009} and then apply Theorem \ref{th:act:stab}. Using Theorem \ref{th:act:det}, we can solve the robust output feedback PDE-ODE stabilization problems considered in \cite{SaGaKr:2018} for a transport equation and a diffusion equation. Finally applying Theorem \ref{th:sen:det}, we can solve the ODE-PDE estimation problem considered in \cite{KrSm:2008} for a transport equation, in \cite{Kri:2009} for a diffusion equation (see also Remark \ref{rm:sen:nonreg}) and in \cite{Kri:2009a} and \cite{LiKr:2010} for a wave equation (see Example 5.2). In the case of Neumann interconnections considered in \cite{SuKr:2010}, we can recover some of the results. We can solve the state feedback PDE-ODE stabilization problem considered in \cite{SuKr:2010} for a wave equation by first stabilizing the wave equation via boundary damping and then using Theorem \ref{th:act:stab}. However, the interconnections in \cite{SuKr:2010} containing heat equations cannot be studied in the framework of this paper because in their formulation as an abstract evolution equation, the control and observation operators are not admissible for the semigroup generated by the state operator. It may be possible to circumvent this admissibility problem by introducing two stable first-order filters in the spirit of Remarks \ref{rm:act:nonreg} and \ref{rm:sen:notcas}. Note that such admissibility problems and transformations like the one used in Example 5.2 are not discussed in the backstepping literature since they implicitly work only with smooth solutions.

We have also presented simple necessary and sufficient conditions for ascertaining the solvability of the stabilization problem for PDE-ODE cascade systems and estimation problem for ODE-PDE cascade systems. To use these conditions, it is enough to find the value of the transfer function of the PDE system at the unstable eigenvalues of the ODE system. The results in this work, unlike the backstepping results, apply to interconnections containing multi-input multi-output systems, higher-dimensional PDEs and PDEs with spatially-varying coefficients.

An important direction for future work is developing an abstract framework, similar to the one in this paper, for studying stabilization problems for coupled PDE-ODE systems. The motivation for this comes from the backstepping works on coupled PDE-ODE systems such as \cite{DeGeKe:18a}, \cite{TaXi:2011}, in which these systems are transformed into PDE-ODE cascade systems. Understanding the transformations they propose in an abstract setting will permit us to develop stabilizing controllers for a class of coupled PDE-ODE systems. Another direction for future research is using the Sylvester equation based approach for adaptive control of PDE-ODE cascade systems. \vspace{-4mm}


\input{references}
\end{document}

%% file: header.tex
\theoremstyle{remark}
\newtheorem{theorem}{Theorem}[section]

\newtheorem{lemma}[theorem]{Lemma}
\newtheorem{assumption}[theorem]{Assumption}

\newtheorem{proposition}[theorem]{Proposition}
\newtheorem{definition}[theorem]{Definition}
\newtheorem{example}[theorem]{Example}

\newtheorem{remark}[theorem]{Remark}

\numberwithin{equation}{section}

\newcommand{\BE}{\begin{equation}}
\newcommand{\rfb}[1]{\mbox{\rm
   (\ref{#1})}\ifx\undefined\stillediting\else:\fbox{$#1$}\fi}

\newfont{\roma}{cmr10 scaled 1200}
\renewcommand{\cline}{{\mathbb C}}

\newcommand{\rline}  {{\mathbb R}}
\newcommand{\sline}  {{\mathbb S}}
\newcommand{\tline}  {{\mathbb T}}

\newcommand{\GGG}  {{\bf G}}

\newcommand{\dd}   {{\rm d}\hbox{\hskip 0.5pt}}
\newcommand{\Ascr} {{\cal A}}
\newcommand{\Bscr} {{\cal B}}
\newcommand{\Cscr} {{\cal C}}
\newcommand{\Dscr} {{\cal D}}
\newcommand{\Escr} {{\cal E}}

\newcommand{\Gscr} {{\cal G}}

\newcommand{\Lscr} {{\cal L}}

\newcommand{\mm}    {{\hbox{\hskip 0.5pt}}}
\newcommand{\m}     {{\hbox{\hskip 1pt}}}

\newcommand{\bluff} {{\hbox{\raise 15pt \hbox{\mm}}}}
\newcommand{\sbluff}{{\hbox{\raise  7pt \hbox{\mm}}}}
\renewcommand{\L}    {{\Lambda}}
\renewcommand{\l}    {{\lambda}}

\newcommand{\FORALL} {{\hbox{$\hskip 11mm \forall \;$}}}

\newcommand{\e}      {{\varepsilon}}

\renewcommand{\o}    {{\omega}}
\renewcommand{\Re}   {{\rm Re\,}}

\newcommand{\bbm}[1]{\left[\begin{matrix} #1 \end{matrix}\right]}
\newcommand{\sbm}[1]{\left[\begin{smallmatrix} #1
   \end{smallmatrix}\right]}